\documentclass[aps,prb,twocolumn,groupedaddress]{revtex4}
\usepackage{epsfig,enumerate,amsmath,amstext,amsthm,amsfonts,color,amssymb}

\def\a{{\alpha}}

\def\b{{\beta}}

\def\lam{{\lambda}}

\def\eps{{\epsilon}}
\def\sig{{\sigma}}
\def\del{{\delta}}

\def\O{{\Omega}}

\def\h{\hbar}

\def\mm{\textbf{m}}
\def\rr{\textbf{r}}

\def\RR{\textbf{R}}
\def\SS{\textbf{S}}
\def\ssig{\overrightarrow{\mathbf{\sigma}}}

\def\kk{\textbf{k}}

\def\qq{\textbf{q}}

\def\pp{\textbf{p}}
\def\PP{\textbf{P}}

\def\cu{\chi^{(1)}}
\def\cd{\chi^{(2)}}
\def\ve{V_{ext}}
\def\dr{\delta\rho}
\begin{document}

\title{Nonlinear response theories and effective pair potentials}
\author{Simon Gravel}
\affiliation{Laboratory of Atomic and Solid-State Physics, Cornell
University, Ithaca, NY, 14850-2501 USA}
\author{N. W. Ashcroft}
\affiliation{Laboratory of Atomic and Solid-State Physics, Cornell
University, Ithaca, NY, 14850-2501 USA}

\date{\today}

\begin{abstract}
We present a general method based on nonlinear response theory to
obtain effective interactions between ions in an interacting
electron gas, which can also be applied to other systems where an
adiabatic separation of time scales is possible. Nonlinear
contributions to the interatomic potential are expressed in terms of
physically meaningful quantities, giving insight into the physical
properties of the system. The method is applied to various test
cases and is found to improve the standard linear and quadratic
response approaches. It also reduces the discrepancies previously
observed between perturbation theory and density functional theory
results for proton-proton pair potentials in metallic environments.
\end{abstract}

\maketitle
\section{Introduction and background}

Perturbation and response theories are ubiquitous in physics. Here
we will discuss response theories as applied to many-component
systems, where the degrees of freedom associated with one component
can be traced out to yield, via an adiabatic separation of time
scales, effective interactions between the remaining components.
 These effective interactions have had their share of success in describing simple
metals \cite{CHW, Haf}, metals with magnetic impurities [i.e. the
Ruderman-Kittel-Kasuya-Yosida (RKKY) interaction \cite{Kit1}], or,
in the classical domain, colloidal suspensions [e.g. the
Derjaguin-Landau-Verwey-Overbeek (DLVO) interaction between charged
colloidal particles \cite{DL,VO} or the depletion interaction
\cite{AO,AO2}]. Linear response in particular has been used
extensively to obtain pair potentials; its simplicity makes it a
very valuable qualitative tool, and in many contexts it also
provides sufficient quantitative accuracy.

Higher-order contributions have been used less often, partly because
calculational complexity increases rapidly with the order of
perturbation considered. The first derivation of the static
quadratic response function for the ground-state noninteracting
electron gas originates with Lloyd and Sholl \cite{LS}. It was later
rederived by Brovman, Kagan \emph{et al}, who wrote a series of
articles applying nonlinear perturbation to metallic systems (see,
for example, Brovman, Kagan, and Kholas\cite{BKK}, and references
therein). A third derivation of the static response function was
provided even later by Milchev and Pickenheim \cite{MP}. A
derivation of the dynamical quadratic response function can also be
found in Pitarke \textit{et al}\cite{PRE}. With the advent of
density functional theory (DFT) and modern computers, response
theory has lost some of its quantitative appeal, but as noted it is
still being used for a qualitative understanding of general trends
in metals \cite{ASR,PRE, lodder:045111}.

Dynamic compression experiments have achieved metallization of
hydrogen at high temperature \cite{WMN}, and there is an expectation
that low-temperature metallization might also be achieved in the
relatively near future \cite{LOL}. Hydrogen has always been
difficult to treat by perturbation approaches because the
interaction of a proton with the electron gas is not weakened by a
repulsive core, as is the case for many elements\cite{LA}. It is not
even obvious that a perturbation approach converges. At high
pressures, though, the enhanced kinetic energy of the electron gas
in hydrogen leads to a better convergence of response theory.
Response-based pair potentials were in fact used to describe with
some success the pressure dependence of the vibron in the hydrogen
solid \cite{NBBA}. On the other hand, the application of standard
DFT methods to hydrogen at very high densities is challenging
because of the strongly inhomogeneous, cusplike electronic density
surrounding the proton and the possible failure of pseudopotentials
designed for a different density range. The pair potentials obtained
from quadratic perturbation theory for dense hydrogen exhibit
discrepancies when compared with ab initio DFT results \cite{BA} for
Wigner-Seitz radii as low as $1.3.$ The Wigner-Seitz radius $r_s$ is
related to the unperturbed electronic density $\rho_0$ by $r_s
a_0=[3/(4 \pi \rho_0)]^{1/3},$ where $a_0=\h^2/m_e e^2$ is the Bohr
radius.

In order to use the perturbation method as a reliable, analytic
alternative approach to DFT in this density range, higher-order
terms are needed. Going beyond quadratic response using conventional
methods is a substantially more difficult task. Accordingly, we
derive a simple identity that allows us to obtain effective pair
potentials beyond quadratic response, which amounts to carrying out
a partial sum of perturbation terms. The result is not only very
intuitive, expressing the pair potential in terms of physically
meaningful expressions, but also very general, since it applies in
any dimension and for both classical and quantum systems. Moreover,
it does not require explicit knowledge of the nonlinear response
functions. It should be noted that if nonlinear effects are to be
fully taken into account, many-body interactions should also be
included, either directly or through effective-medium approaches
\cite{JNP}. This article focuses on obtaining pair potentials that
can be used as a starting point for either approach.

In Section \ref{secresp}, we outline the response theory formalism
and derive the simplest version of our result, equation \ref{resum}.
In Section \ref{interact} (and Appendix I) a generalization is
derived that applies to homogeneous, interacting systems. The
intuitive nature of this result permits us to explain part of the
discrepancies observed between the results of Nagao \textit{et al}
\cite{NBBA} and Bonev and Ashcroft\cite{BA}, and to improve upon
quadratic response. In Section \ref{future} we apply our results to
obtain pair potentials in various systems, including metallic
hydrogen. Generalizations to many-center interactions and magnetic
perturbations are discussed in Appendix II.

\section{Effective interactions and response theory}\label{secresp}

 We start with a canonical system composed of a mixture of at least two different types of
particles, in contact with a heat bath establishing a temperature
$T.$ The ground-state properties of the system can be studied from
the limit $T\rightarrow 0.$

The particles will be divided in two subsets, labeled $L$ (for
Light, typically electrons) and $M$ (for Massive, typically ions).
It is assumed that all $L$ particles are identical. On the other
hand, the set $M$ can contain various types of particles. The time
scales associated with the $L$ particles will be assumed to be much
shorter than those associated with the $M$ particles, allowing the
use of an adiabatic separation of time scales. In real systems, this
separation is not exact. The assumption of exact separation of time
scales, which will be made throughout this article, is referred to
as the adiabatic or Born-Oppenheimer approximation. We will not
discuss the accuracy of this approximation here and refer the reader
to the existing literature (see, e.g., Ziman\cite{Zim}).

The adiabatic approximation allows the treatment of the degrees of
freedom associated with the light particles $L$ in an average way,
leading to a much simpler \emph{effective} Hamiltonian for the
remaining $M$ particles.

The Hamiltonian of the initial system has the form

\begin{equation}\label{Ham}
\begin{split}
    H=&\sum_{i} \frac{\pp^{2}_i}{2 m}+\sum_{j}
\frac{\PP^{2}_j}{2M_j}+V^M(\{\RR\})+V^L(\{\rr\})\\&+V^{LM}(\{\rr\},\{\RR\}).
\end{split}
\end{equation}

The notation $\{\rr\}$ designates the set of all coordinates
$\{\rr_i\}_{i=1,...,N_L},$ and $\pp_i,\rr_i,$ and $m$ are the
momenta, positions and mass of $L$ particles, while $\PP_i,\RR_i,$
and $M_i$ are the corresponding quantities for $M$ particles. We
imagine the system to be confined to a macroscopic volume $\Omega.$

In many relevant physical systems the interactions between the
particles are largely pairwise. An important example of such a
system is a metal, where the $L$ particles could be the valence
electrons and the $M$ particles the ionic cores. For pairwise
systems we can write

\begin{equation}\label{pairwise}
\begin{split}
 V^L(\{\rr\})&=\sum_i\phi_L^{(1)}(\rr_i)+\sum_{i\neq
j} v^L{(\rr_i,\rr_j)},\\
V^M(\{\RR\})&=\sum_i\phi_M^{(1)}(\RR_i)+\sum_{i\neq
j} v^M_{ij}{(\RR_i,\RR_j)},\\
V^{LM}(\{\rr\},\{\RR\})&=\sum_{ij} v^{LM}_j{(\rr_i,\RR_j)},\\
\end{split}
\end{equation}

\noindent where $\phi_L^{(1)}$ and $\phi_M^{(1)}$ denote external,
one-particle potentials, $v^L$ is the pair potential between $L$
particles, $v^M_{ij}$ is the pair potential between $M$ particles
$i$ and $j,$ and $v^{LM}_j$ is the pair potential between the $M$
particle $j$ and the $L$ particles. In the case of long-range,
Coulombic interactions, care should be taken in the definition of
the pair potentials to ensure that both the unperturbed and the
perturbing systems are thermodynamically well
defined\cite{commentaire}.

We will restrict our attention to such pairwise systems, even though
our main results require only the slightly weaker assumptions that
we can write the interaction between $L$ and $M$ particles as

$$V^{LM}(\{\rr\},\{\RR\})=\sum_{i} V^{LM}_1(\rr_i,\{\RR\}).$$

This is to say that the joint effect of all the $M$ particles, if
they were held fixed, would amount to an additional one-body
potential for the $L$ particles.

\subsection{Effective interactions and the adiabatic separation of time scales}

As mentioned, we want to find an \textit{effective} Hamiltonian for
the $M$ particles by tracing out the degrees of freedom associated
with the $L$ particles. Since volume $(\O$) and temperature ($T$)
are specified, we begin with the relevant Helmholtz free energy $F,$

\begin{equation}\label{trace1}
F=-kT \ln\textrm{Tr}_{\{LM\}} e^{-\b H}.
\end{equation}

We then trace over the degrees of freedom associated with $L$ at
fixed configuration of $M,$ assuming an adiabatic separation of
time-scales,

\begin{equation*}
F=-kT \ln\left(\textrm{Tr}_{M} e^{-\b (T^{M}+V^M)}
\textrm{Tr}_{L(M)} e^{-\b (T^{L}+V^L+V^{LM})}\right),
\end{equation*}

\noindent where $\textrm{Tr}_{L(M)}$ means the trace over states of
particles of type $L$ for a fixed configuration of particles $M,$
and $T^{L,M}$ is the kinetic energy associated with the L and M
particles, respectively. The free energy of system $L$ for a fixed
configuration of $M$ is simply
\begin{equation}\label{trace3}
     F_L(\{\RR\},T)= -kT \ln\left(\textrm{Tr}_{L(M)} e^{-\b
(T^{L}+V^L+V^{LM})}\right).\end{equation}

The total free energy can then be written as

\begin{equation}\label{trace2}
F=-kT \ln\textrm{Tr}_{M} e^{-\b H^{eff}_M},
\end{equation}

\noindent where the effective Hamiltonian has the form
\begin{equation*}
    H^{eff}=\sum_{j}
\frac{\PP^{2}_j}{2M_j}+V^M(\{\RR\})+V^{eff}(\{\RR\})
\end{equation*}

\noindent and $V^{eff}(\{\RR\})=F_L(\{\RR\},T)$ is the desired
effective interaction between particles of type $M$. It is clearly
state dependent, since it depends on temperature. It also depends on
the volume $\O$ and on the properties of the $L$ particles,
including their mass, their number, and the form of their
interactions. Note that even if the initial Hamiltonian contains
only pairwise interactions, the \textit{effective} Hamiltonian will
typically contain many-center interactions as well as
volume-dependent but structure-independent terms. Here we will focus
on the determination of pair interactions.

In order to obtain the effective interaction we need to calculate
the free energy \eqref{trace3} of system $L$ while the $M$ particles
are held fixed. To do this we treat the potential
$V_1^{LM}(\rr,\{\RR\})$ as an external one-body perturbation to the
system composed of particles of type $L.$ In the following
$V_1^{LM}(\rr,\{\RR\})$ will therefore be simply written as
$V_{ext}(\rr),$ keeping in mind the dependence of $V_{ext}$ on the
$\RR_i$. We then proceed to a functional expansion of the free
energy in orders of $V_{ext}(\kk)$\cite{commentaire}
$$V_{ext}(\kk)=\int_\Omega d\rr e^{-i\kk\cdot\rr} V_{ext}(\rr).  $$

\subsection{Response theory, and the coupling constant integration method}

 The change in density induced in a system by a perturbing external potential $V_{ext}(\kk)$ takes the form

\begin{widetext}
\begin{equation}\label{respdef}
\begin{split}
  \delta\rho(\kk,T)&=\rho(\kk,T)-\rho^{(0)}(\kk,T)\\&=\sum_{\kk'}\cu(-\kk,\kk',T) \ve(\kk')+\frac{1}{\O}\sum_{\kk',\kk''}
  \cd(-\kk,\kk',\kk'',T) \ve(\kk')\ve(\kk'')+\cdots
\end{split}
\end{equation}
\end{widetext}

The functions $\chi^{(n)}$ are, by definition, the response
functions of the unperturbed system (which at this point is not
necessarily uniform) and carry all the information about this
system, including temperature. We have assumed a large but finite
volume $\O$ and a dense but discrete distribution of wavevectors
$\kk,$ since our main interest is the electron gas. Continuous
systems can be recovered using the usual prescription
\cite{commentaire} [in three dimensions, it reads $1/\O \sum_\kk
\rightarrow \int d\kk/(2\pi)^3$].
 Even though the response functions depend on temperature, the
 methods we present here do not involve this temperature dependence explicitly. In order to
 simplify the notation, we will not explicitly keep track of the temperature in the
following. Unless otherwise specified, $\cu(\kk,\kk')$ will simply
stand for $\cu(\kk,\kk',T),$ etc.

Equation \eqref{respdef} can be used together with the
coupling-constant integration method \cite{commentHL,PN} to obtain
the variation in the Helmholtz free energy arising from the
perturbation $ V_{ext}:$ namely,

\begin{equation}\label{coupint}\Delta F=\int_0^1 d\lam
\left<V_{ext}\right>_\lam,\end{equation}
 \noindent
where $\left<\cdot\right>_\lam$ is the statistical average with
respect to the states of the Hamiltonian $H_\lam^{ext},$ which is
obtained by replacing $V_{ext}$ by $\lam V_{ext}$ in $H^{ext}$. In
the thermodynamic integration scheme \cite{elber1996rdt}, the
integrand in equation \eqref{coupint} is determined by numerical
simulation. In the perturbation approach, it is instead expanded in
powers of the external perturbation, yielding
\begin{widetext}
\begin{equation}\label{fgen}
\begin{split}
\Delta &F =\frac{1}{\O}\sum_{\kk,n} \frac{\rho^{(n)}(\kk)
V_{ext}(-\kk)}{n+1} =\sum_{n=0}^\infty \frac{1}{(n+1)\O^n}
\sum_{\kk_1,\ldots,\kk_{n+1}} \chi^{(n)}(\kk_1,...,\kk_{n+1})
V_{ext}(\kk_1)... V_{ext}(\kk_{n+1}),
\end{split}
\end{equation}
\end{widetext}
\noindent where $\rho^{(n)}$ is the part of \eqref{respdef} that is
of order $n$ in $V_{ext}.$ The "zeroth order" response function is
related to the density of the unperturbed system,
$\chi^{(0)}(\kk)=\rho^{(0)}(-\kk)/\O.$

Note that terms of $n^\text{th}$ order in the external potential in
expression \eqref{respdef} yield terms of order $n+1$ in
\eqref{fgen}. In the following, "$n^\text{th}$ order response"
refers to terms of order $n$ in \eqref{respdef}, unless otherwise
specified.

In order to study pair potentials, let us now assume that the
perturbation originates with \emph{two} external sources, located at
positions $\RR_a$ and $\RR_b,$ so that

$$V_{ext}(\kk)=V_a(\kk)e^{i\kk\cdot\RR_a}+V_b(\kk) e^{i\kk\cdot\RR_b}. $$

The free energy now depends on $\RR_a$ and $\RR_b$ (and would depend
only on $\RR_a-\RR_b$ if we had restricted ourselves to initially
homogeneous systems). This yields an induced effective pair
potential $\phi^{i}(\RR_a,\RR_b)$ between the two sources, defined
as the sum of all terms in \eqref{fgen} that depend on both $V_a$
and $V_b$. Here the standard approach is to keep only terms up to a
given order in $V_{ext}.$ Linear response yields the \emph{induced}
pair potential

\begin{equation}\label{lin}
\begin{split}
\phi^i_{lin}(\RR_a,\RR_b)=\frac{1}{\O} \sum_{\kk,\kk'}&
\cu(\kk,\kk')\\&\times V_a(\kk) V_b(\kk')e^{i(\kk
\cdot\RR_a+\kk'\cdot\RR_b)}.
\end{split}
\end{equation}

In order to emphasize the role of the coupling-constant integration
and to hint at an upcoming result [equation \eqref{resum}], this can
be written as
\begin{widetext}
\begin{equation}\label{linpref}
\begin{split}
\phi^i_{lin}(\RR_a,\RR_b)=&\frac{1}{\O} \sum_{\kk} \left(V_a(\kk)
\delta \rho^{lin}_b(-\kk,\RR_b)e^{i\kk \cdot \RR_a}+V_b(\kk) \delta
\rho^{lin}_a(-\kk,\RR_a)e^{i\kk\cdot
\RR_b}\right)\\&-\frac{1}{\O}\sum_{\kk_1,\kk_2}
\chi^{(1)}(\kk_1,\kk_2) V_{a}(\kk_1) V_{b}(\kk_2) e^{i
(\kk_1\cdot\RR_a+\kk_2\cdot \RR_b)},
\end{split}
\end{equation}
\end{widetext}

\noindent where $\delta\rho^{lin}_i(-\kk,\RR_i)$ is the density
induced, at linear order, by particle $i$ located at position
$\RR_i.$

The first two terms in this expression correspond to the Coulombic
energy of the system at the level of linear response. The third
term, which arises from the coupling-constant integration,
incorporates the variation in kinetic energy and entropy caused by
the perturbation.

Instead of keeping terms linear in $V_{ext},$ we can choose to keep
all terms that are linear in $V_a,$ for all orders in $V_b.$ In this
case we find, using equation \eqref{respdef}, that the pair
potential can be written as

\begin{equation}\label{flin}
\begin{split}
\phi^i(\RR_a,\RR_b)&\simeq \phi_{SPb}^i(\RR_a,\RR_b)\\&=
\frac{1}{\O} \sum_{\kk} V_a(\kk) \delta\rho_b(-\kk, \RR_b)e^{i\kk
\cdot\RR_a}.
\end{split}
\end{equation}

Here, $\delta\rho_b(\kk,\RR_b)$ is the total density that would be
induced if the perturbation potential was caused by source $b$
alone, i.e.,  $V_{ext}(\kk)=V_b(\kk)e^{i\kk \cdot \RR_b}.$ Notice
that the terms arising from the interaction between $\delta \rho_a$
and $V_b$ and the change in kinetic energy and entropy arising from
the perturbation (taken into account by the coupling-constant
integration) cancel each other out exactly in this case. One can
also obtain a different estimate
 to the potential by inverting the roles of $a$ and $b$ in the previous discussion, leading to
 $\phi_{SPa}^i.$

This asymmetric approach will be referred to as the successive
perturbation method, or SPM. Indeed, one interpretation of this
result is that instead of perturbing the initial system with $a$ and
$b$ simultaneously, we initially add only $b,$ determine the
properties of the intermediate system, and then, subsequently, add
particle $a.$ Since our pair potentials result from the calculation
of a free energy, the Gibbs-Bogoliubov inequality applies and
equation \eqref{flin} is related to a rigorous bound on the pair
potential: namely,

$$\phi^i(\RR_a,\RR_b)\leq \phi_{SPb}^i(\RR_a,\RR_b)+\langle V_a \rangle_0-\Delta F_a,$$

\noindent where $\langle V_a \rangle_0$ is the statistical average
of the operator $V_a(\kk)e^{i\kk \cdot \RR_a}$ with respect to
states of the initial, unperturbed system, and $\Delta F_a$ is the
change in free energy induced by adding $a$ to the unperturbed
system, which can be obtained by setting $V_b$=0 in \eqref{fgen}.

Equation \eqref{flin} is the first term in the expansion of the
energy if $b$ is treated exactly and $a$ is treated as a
perturbation.  Note that higher-order terms in the expansion would
simply involve higher-order functional derivatives of the free
energy of the intermediate system including $b$, which are related
to density-density correlation functions of that system.

 The SPM is especially useful if $V_a$ is weak and
$V_b$ is strong \cite{pscp}, but is not as useful when both sources
of perturbation are strong and require nonlinear treatment. On the
other hand, we can easily obtain from this relation another
expression for the induced pair potential that incorporates all
contributions that are linear in \emph{either} source. This
expression therefore includes all contributions to the pair
potential up to cubic order in $V_a$ and $V_b.$ It reads
\begin{widetext}
\begin{equation}\label{resum}
\begin{split}
\phi^i_{SSP}(\RR_a,\RR_b)=&\frac{1}{\O} \sum_{\kk} \left(V_a(\kk)
\delta \rho_b(-\kk,\RR_b)e^{i\kk \cdot \RR_a}+V_b(\kk) \delta
\rho_a(-\kk,\RR_a)e^{i\kk\cdot
\RR_b}\right)\\&-\frac{1}{\O}\sum_{\kk_1,\kk_2}
\chi^{(1)}(\kk_1,\kk_2) V_{a}(\kk_1) V_{b}(\kk_2) e^{i
(\kk_1\cdot\RR_a+\kk_2\cdot\RR_b)}.
\end{split}
\end{equation}
\end{widetext}

This result, which emerges from what we will refer to as the
symmetrized successive perturbation method (SSPM), can also be seen
to be a natural generalization of linear response \eqref{linpref}.
Equation \eqref{resum} is our main result for noninteracting
systems. It also applies to interacting systems, but in that case it
can be improved upon. We will do this in Section \ref{interact}.

Equation \eqref{resum}, despite its simplicity, has many interesting
features:

(a) It is very general; it can be used in any dimension, for
classical and quantum, homogeneous and inhomogeneous, noninteracting
and interacting systems (although, as mentioned, it can be improved
for interacting systems; see Section \ref{interact}). (b) It is
intuitive: the first two terms are the interaction of the potential
energy associated with each perturbation with the density induced by
the other. The last term, which is equal to the negative of the
linear response potential, accounts for the change in kinetic energy
and entropy of the perturbed system and the contributions to the
density that are not additive in $V_a$ and $V_b.$ (c) It includes
all contributing terms up to quadratic response [yielding cubic
terms in equation \eqref{fgen}], plus 8 out of 14 contributing terms
of cubic response: it includes more terms than quadratic response.
(d) It does not require the explicit knowledge of the second-order
response function, but only that of the linear response function of
the initial, unperturbed system. (e) It expresses the pair
potentials in terms of quantities that are simple, symmetric, and
can in principle be measured. Finally (f), since it takes the effect
of a single, isolated perturbation as an external input, equation
\eqref{resum} allows us to treat the effect of stronger, localized
perturbations which are often difficult to treat with purely
perturbative methods.

We use this result in two test cases in section \ref{future}. The
first example is the effective interaction between particles
perturbing, via delta-function potentials, a noninteracting quantum
one-dimensional electron gas. The second is a version of the
classical Asakura-Oosawa model \cite{AO,AO2} of the depletion
interaction, with finite square wells replacing hard-sphere
potentials. We then apply it to the more realistic calculation of
the pair potential between protons in a metallic environment.

Before we do this, we use the intuitive form of equation
\eqref{resum} to suggest the existence of a higher-order correction
that applies to interacting systems, which will be derived in
Appendix I.

\section{Corrections specific to interacting systems}
\label{interact}

 As mentioned, equation \eqref{resum} is valid for the interacting
electron gas as well as for the noninteracting one: the interactions
simply modify the forms of the $\chi^{(n)}.$ But upon further
inspection of this equation, one might wonder about the absence of
coupling between the induced densities themselves. The part of this
interaction that is linear in $V_a$ or $V_b$ is included in
\eqref{resum}, but the part that is nonlinear in both $V_a$ and
$V_b$ is not. It turns out that some of these contributions can also
be expressed intuitively in terms of $\delta\rho_a$ and
$\delta\rho_b.$

For \emph{homogeneous} interacting systems we obtain, by summing up
higher-order terms that correspond to reducible diagrams (see
Appendix I), an extra term of the form

\begin{equation}\label{elel}
   \phi_{red}(\RR_{ab})= \frac{1}{\O} \sum_\kk\delta\rho^{NL}_a(\kk)\tilde
   v(k)\delta\rho^{NL}_b(-\kk) e^{i \kk\cdot \RR_{ab}},
\end{equation}

\noindent where $\RR_{ab}$ is the distance between $a$ and $b$,
$\delta\rho^{NL}_a=\delta\rho_a-\delta\rho^{(1)}_a$, and $\tilde v$
is the effective interaction between the induced densities. It reads

\begin{equation}\label{vtilde}
    \tilde v(k)=\eps(k) c_0^{(1)}(k),
\end{equation}

\noindent where
$$\eps(k)=1-c_0^{(1)}(k)\chi_0(k)$$

\noindent involves the \emph{noninteracting} response function
$\chi_0^{(1)}.$ In classical statistical mechanics $c_0^{(1)}(k)$ is
the Ornstein-Zernike function for the unperturbed system, multiplied
by $k_BT$. In the case of quantum mechanical particles of charge $e,$ we have\cite{commentaire}
\begin{equation*}
  c_0^{(1)}(k)=v_c(k)+\mu_1(k),
\end{equation*}

\noindent where $v_c(k)=4 \pi e^2/k^2$ is the Coulomb interaction
and $\mu_1$ is the first functional derivative with respect to
density of the exchange-correlation potential. It is related to the
local-field correction $G$ by $\mu_1(k)=-G(k) v_c(k).$

With some reorganization, we can now write the total induced pair
potential in a remarkably simple form: namely,
\begin{widetext}
\begin{equation}\label{final}
    \phi^i_{SSP}(\kk)=\eps(k)\left[V_a(\kk) \dr_b(-\kk)+\dr_a(\kk) V_b(-\kk)
+\dr_a(\kk) c_0^{(1)}(k)\dr_b(\kk)-V_a(\kk)\chi_0^{(1)}(k)
V_b(-\kk)\right].
\end{equation}
\end{widetext}
In real space,

 $$\phi^i(\RR_{ab})=\frac{1}{\O} \sum_\kk \phi^i(\kk) e^{i \kk \cdot \RR_{ab}}.$$

This result brings up an important point. The energy associated with
the interaction between potential $V_b$ and the $n^\mathrm{th}$
order contribution to the density induced by $a$, which we write as
$\delta\rho_a^{(n)},$ arises at order $n+1$ in the energy expansion.
On the other hand, the "electron-electron" interaction between
$\delta\rho_a^{(n)}$ and $\delta\rho_b^{(n)}$ appears only at order
$2 n$ in the energy.
 Except for the special case of linear response, the
electron-electron interaction terms originate with a higher order in
response than the corresponding electron-perturbation terms.
Therefore termination of the series \eqref{fgen} at any order beyond
linear response will typically result in pair potentials that are
overly attractive.

Indeed, if we compare the effective pair potentials for hydrogen
atoms in jellium obtained from quadratic response \cite{NBBA} to
those obtained from ab initio methods \cite{BA}, we observe exactly
such a discrepancy. We calculate in section \ref{Hydrogen} the
lowest-order contribution arising from \eqref{elel}, and find that
it indeed improves the agreement between ab initio methods and
response theory.

We finally draw the reader's attention to the similarity (for
protons in an electron gas) between the ab initio equation
\eqref{final} and the variational Heitler-London evaluation of the
isolated hydrogen molecule energy. This will be further discussed in
section \ref{Hydrogen}.

\section{Examples of applications} \label{future}

\subsection{Delta functions in a noninteracting electron gas} \label{deltas}
As a simple instructive example, we first consider a one-dimensional
noninteracting electron gas of unperturbed linear density $\rho_0$
confined to a large length $L$, with periodic boundary conditions.
The external perturbations have the form

$$V_{\a}(r)=\frac{\h ^2 u}{2 m} \delta(r)~~~~~~~~~~~~~~~(\a=a,b).$$

From \eqref{resum} we obtain immediately

\begin{equation}\label{delta1}
\phi_{SSP}(R_{ab})\simeq \frac{\h ^2 u}{ 2 m} (2 \delta
\rho_a(R_{ab})- \delta \rho^{lin}(R_{ab})).
\end{equation}

Here

\begin{equation}\label{deltlin}
\delta \rho^{lin}(r)= \frac{\h ^2 u}{ 2 m} \sum_k \chi^{(1)}(k,-k)
e^{i k r}
\end{equation}

\noindent and $\chi^{(1)}$ is the linear response function of the
one-dimensional noninteracting electron gas \cite{Yaf, Kit1}:

$$\chi^{(1)}(k_1,k_2)=\frac{2 m}{\h^2 \pi
k_1} \ln\left|\frac{k_1+2k_F}{k_1-2 k_F}\right| \delta_{k_1,-k_2}.$$

The Fermi wave-number $k_F$ is related to the unperturbed linear
density by $k_F=\pi\rho_0.$

Converting the sum into an integral, we find (see Kittel
\cite{Kit1}, and also Yafet \cite{Yaf} and Giuliani \textit{et
al}\cite{GVD})

\begin{equation}\label{dellindelt}
\delta \rho^{lin}(r)= \frac{u}{\pi} \mathrm{Si}(2 k_F
r)-\frac{u}{2},
\end{equation}

\noindent where $\mathrm{Si}$ is the sine integral function
\cite{abramowitz1966hmf}. The nonlinear induced density $\delta
\rho_a(r)$ can also be calculated exactly as the sum of the bound
and scattering state density: namely,

$$\delta \rho_a=\delta \rho_{bound}+\delta \rho_{s},$$

\noindent with

\begin{equation}\label{bound}
    \delta\rho_{bound}(r)= -\Theta(-u) u e^{u|r|}=\left\{\begin{array}{ll}
                                 -u e^{u|r|} &\mbox{if $u<0$},\\
                                 0 &\mbox{otherwise}.
                                 \end{array}\right.
\end{equation}

Here $\Theta$ is the Heaviside step function.

The scattering state density is, in the limit of large $L,$
\cite{GVD}

\begin{equation}\label{scatt}
    \delta\rho_{s}(r)=\frac{2}{\pi}\int_0^{k_F}dk\left(\frac{2uk \sin(2 k |r|)}{4 k^2+u^2}-\frac{u^2\cos(2 k |r|)}{4
k^2+u^2}\right).
\end{equation}
Factors of $2$ are included for spin degeneracy. This integral can
be evaluated in terms of the exponential integral $E_1$ using, for
example, relations 5.1.41 and 5.1.42 from Abramowitz and Stegun
\cite{abramowitz1966hmf}. We find

\begin{equation}\label{scatt2}
  \delta\rho_{s}(r)=\frac{e^{u |r|} u}{\pi}\left(\Im m\left[\mathrm{E_1}\left(\left(u +2 i k_f\right)
|r|\right)\right]+\pi\Theta(-u)\right).
\end{equation}

The total density induced by a single delta function potential in a
noninteracting electron gas therefore takes the very simple form

\begin{equation}\label{delrhodelt}
  \delta\rho_a(r)=\frac{u e^{u |r|} }{\pi}\Im m\left[\mathrm{E_1}\left(\left(u +2 i k_f\right)
|r|\right)\right].
\end{equation}

Note that for $k_f\rightarrow 0$ we find that
$\delta\rho=\dr_{bound}$, as expected. The appearance of a bound
state at $u=0$ corresponds to the branch cut of $\mathrm{E}_1$ along
the negative real axis.

If we use \eqref{dellindelt} and \eqref{delrhodelt} in
\eqref{delta1}, we find an expression for the pair potential as a
function of $R_{ab}$ which reads

\begin{equation}\label{phidel}
\begin{split}
  \phi_{SSP}(R_{ab})&=\frac{\h^2 u^2}{ \pi m} e^{u R_{ab}} \left(\Im m\left[\mathrm{E_1}\left(\left(u +2 i k_f\right) R_{ab}\right)\right]\right)\\
 &-\frac{\h^2 u^2}{2 \pi m} \left(\mathrm{Si}(2 k_F R_{ab})-\frac{\pi}{2}\right).
\end{split}
\end{equation}

Alternatively, the \emph{exact} pair potential can be calculated by
solving the Schr\"odinger's equation directly for different values
of $R_{ab}$. The result is expressed as a sum over the eigenvalues
of the Hamiltonian, which are obtained by numerically solving a set
of transcendental equations.

\begin{figure}
\scalebox{.3}{\includegraphics{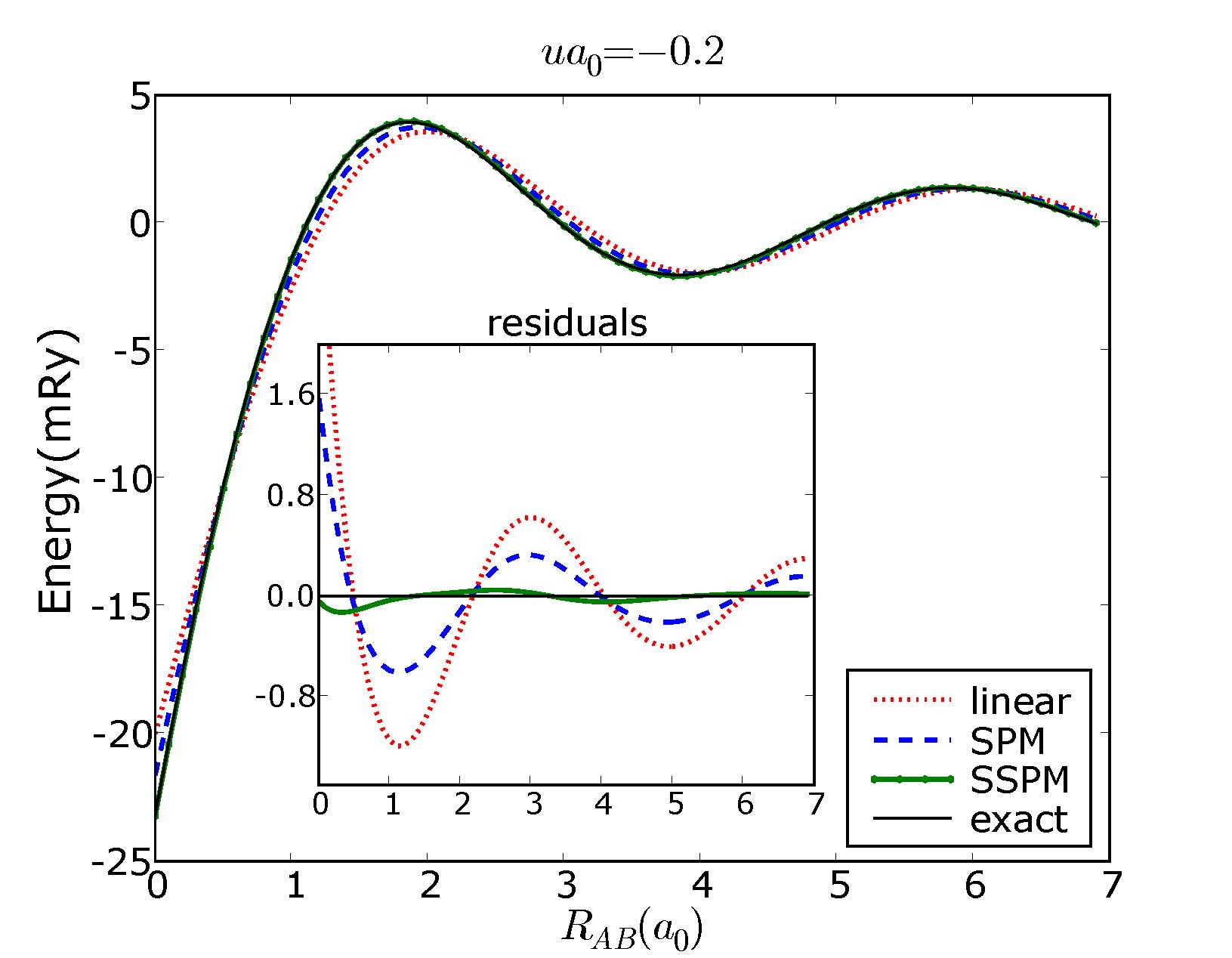}}
\scalebox{.3}{\includegraphics{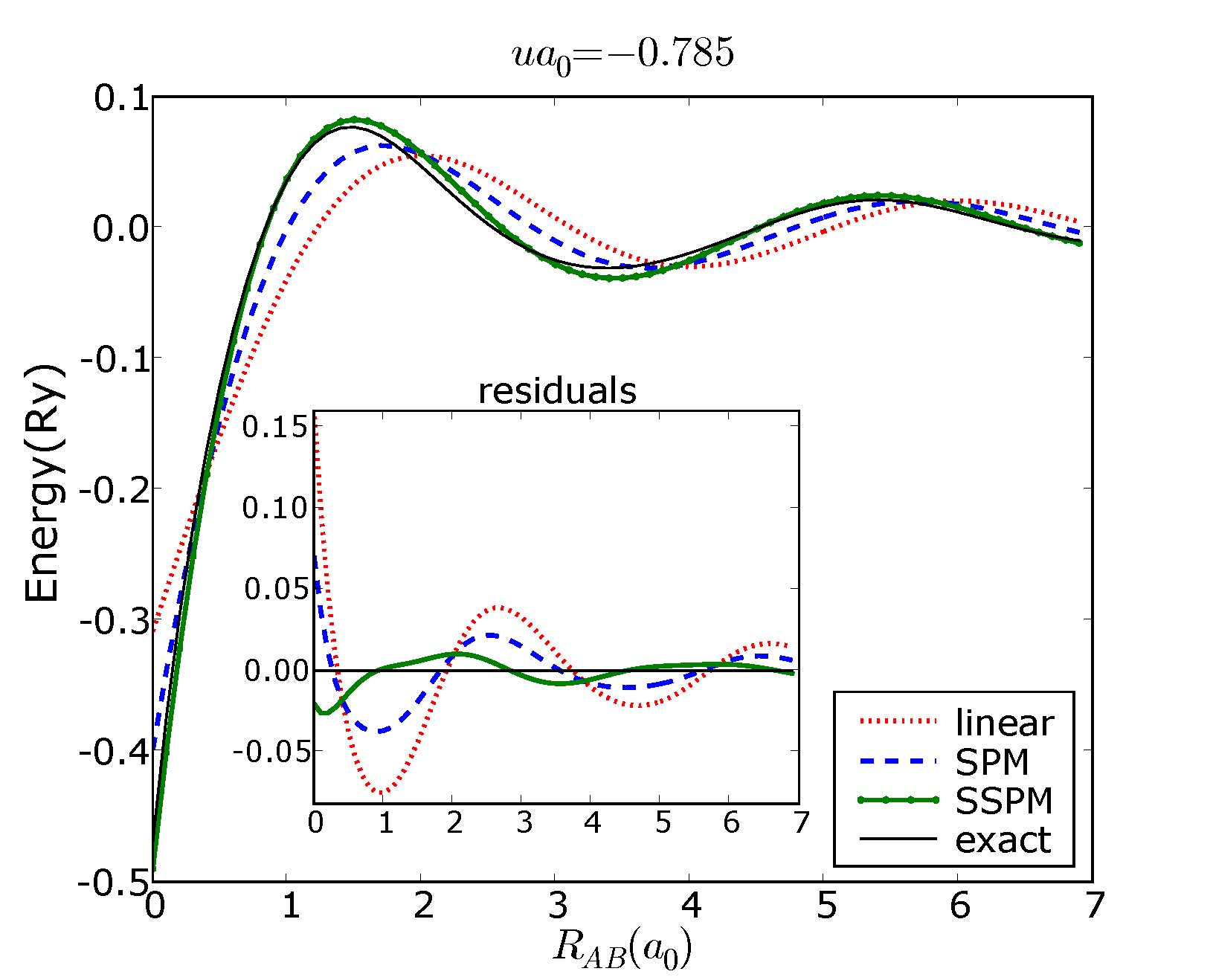}}\\
\scalebox{.3}{\includegraphics{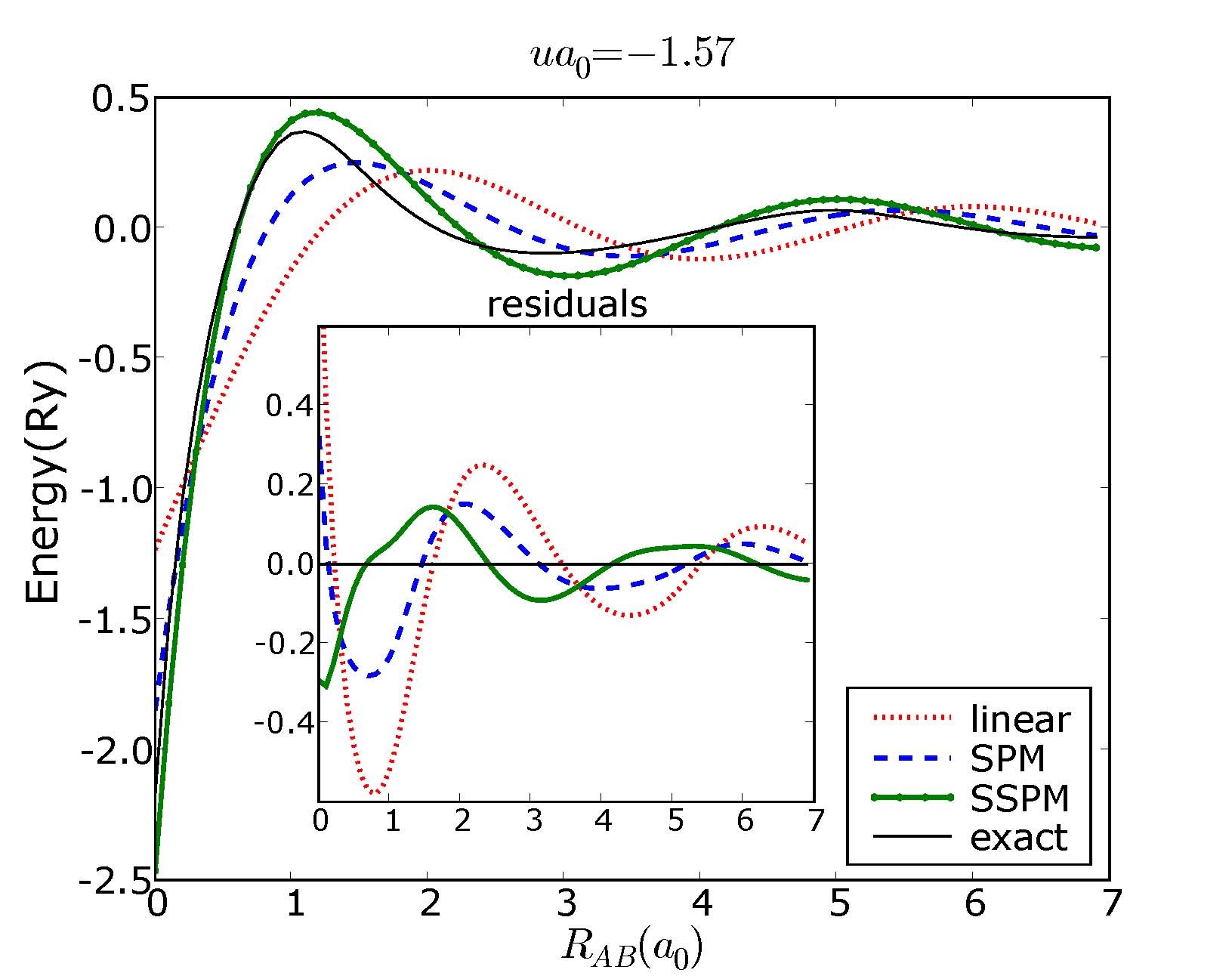}}
\scalebox{.3}{\includegraphics{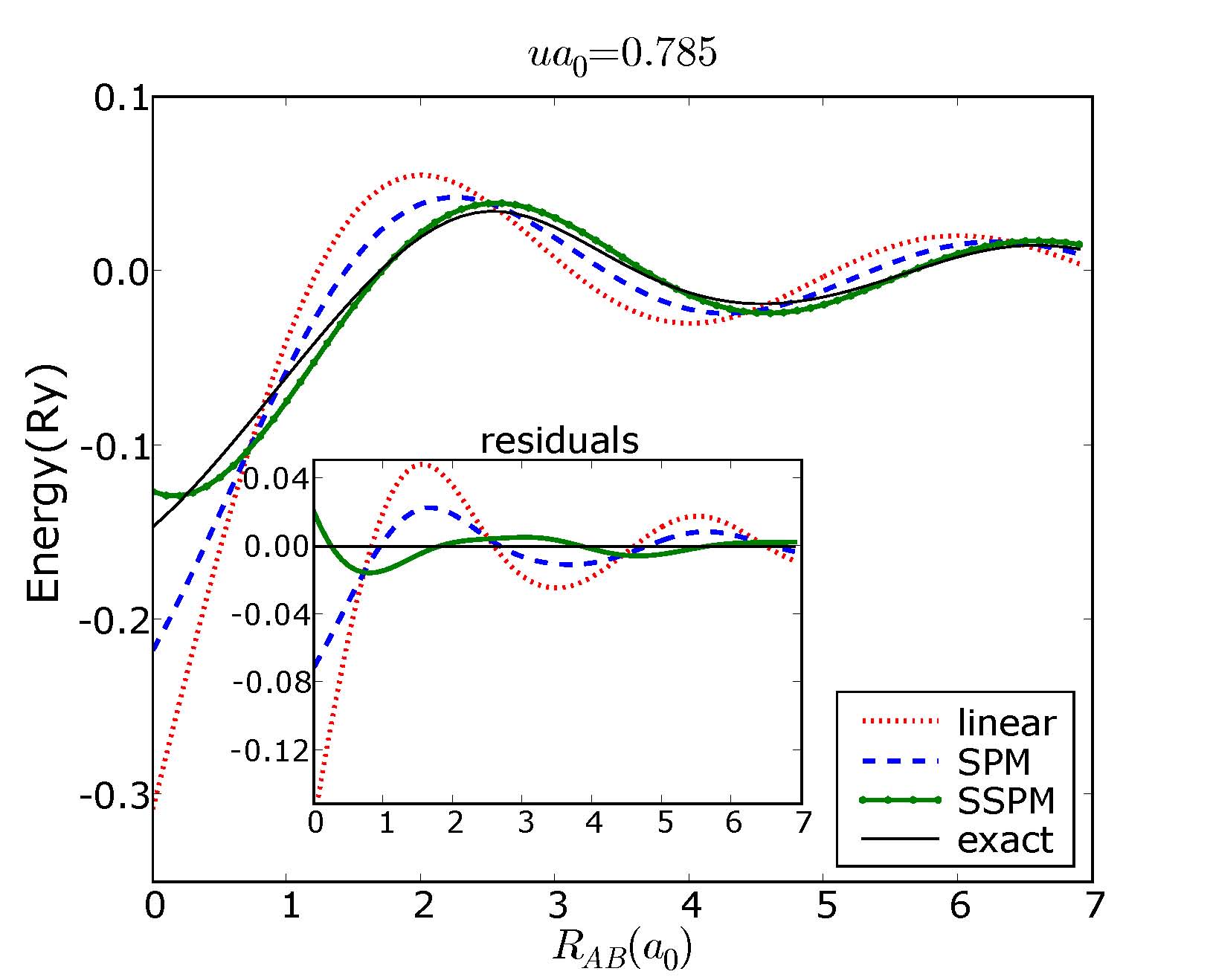}}

 \caption{(color online) Pair potentials for delta function potentials in one dimensional noninteracting electron gas:
exact, linear response, and the SPM and SSPM results. The Fermi
wavevector is set to $k_f=0.785/a_0.$ The residuals (difference
between estimated and exact pair potentials) are shown in the inset.
Note the difference in behavior of the SSPM near the origin between
attractive (first three figures) and repulsive (last figure) delta
function potentials.\label{ppdeltas}}
\end{figure}

We compare in Figure \ref{ppdeltas} these numerical pair potentials
with the analytical ones obtained from the SPM and SSPM [equation
\eqref{phidel}], and those obtained from linear response, for
various interaction strengths, including attractive and repulsive
cases. We observe that the SSPM improves upon linear response and
the SPM, especially quantitatively, at low $u,$ but also
qualitatively (at higher $u.$)

As a side remark, notice that the system has no bound state for
$u>0,$ one even bound state for $u<0$, and one extra odd bound state
when $R_{ab}u<-2$. It has been suggested \cite{Louis} that the
appearance of a bound state might cause the failure of response
theory, since such states are qualitatively different from the
initial, unperturbed free-electron system. We see that this is not
the case in this particular example and that the pair potentials can
be accurately described by response theory despite the presence of a
bound state. This is consistent with the observation that no
discontinuity in the density of charge occurs upon the formation of
a localized, bound state (see, e.g., Galindo  \textit{et
al}\cite{galindo2005cte}).

\subsection{Classical depletion interaction}
The depletion interaction (or entropic attraction) plays a major
role in classical colloidal systems, where it is used to tune the
interaction between colloidal particles immersed in a solvent. The
addition of polymers to the solvent indeed causes an effective
attraction between the colloidal particles which can be adjusted by
modifying the polymer concentration.

 The Asakura-Oosawa model \cite{AO,AO2} describes this effect by representing
 colloidal particles as hard spheres of radius $D$ and (folded) polymers as hard
spheres of radius $\delta$. The interactions between the polymers
are neglected. The polymers are therefore treated as an ideal gas
which is excluded from spheres of radius $D+\delta$ surrounding each
colloid. Here we will replace the hard-sphere potential with a
finite repulsion of magnitude $V_0.$ Within this model the effective
attractive potential between the colloids can be calculated exactly
and compared to the response theory results, providing a useful
benchmark for perturbative approaches in the classical regime.

Suppose we have only two colloidal particles (at positions
$\RR_{a,b}$) in a bath of polymers of unperturbed density
$\rho_0=N/\O$. Since the polymers are taken not to interact, their
density is given by

\begin{equation}\label{rhocoll0}
    \rho(\rr)=\frac{N e^{-\b V_e(\rr)}}{\int_\O d\rr' e^{-\b V_e(\rr')}
    },
\end{equation}

\noindent where $V_e(\rr)=V_a(\rr)+V_b(\rr)$ is the total potential
and
$$V_{a,b}=\left\{\begin{array}{ll}V_0 & \mbox{if $|\rr-\RR_{a,b}|<D+\delta,$
}\\0&\mbox{otherwise}.\end{array}\right.$$

Since $V_e(\rr)$ is zero except within a bounded region, $\rho(\rr)$
simplifies, in the thermodynamic limit, to

\begin{equation}\label{rhocoll}
    \rho(\rr)=\rho_0 e^{-\b V_e(\rr)}.
\end{equation}

Now we can use the coupling-constant integration method to obtain
the Helmholtz free energy:

\begin{equation}\label{encoll}
\begin{split}
    F-F_0=&\int_0^1d\lam \int d\rr V_{ext}(\rr) \rho_\lam(\rr)\\
=&- A_o(\RR_a,\RR_b) \frac{\rho_0}{\b} \left(e^{-2 \b
V_0}-1\right)\\
 &-A_{r}(\RR_a,\RR_b) \frac{\rho_0}{\beta} \left(e^{- \b V_0}-1\right),
\end{split}
\end{equation}

\noindent where $A_o$ is the overlap volume between the two spheres
of radius $D+\delta$ centered at $\RR_{a,b}$, and $A_{r}=8 \pi
(D+\delta)^3/3-2 A_o$ is the non-overlapping volume of the spheres.
All the $\RR_{a,b}$ dependence is contained in
$A_o(\RR_a,\RR_b)=A_o(\RR_a-\RR_b).$ Accordingly we write the pair
potential as

\begin{equation}\label{phi}
\phi^i(\RR_a-\RR_b,V_0)    = -\frac{\rho_0 A_0(\RR_a-\RR_b)}{\b}
\left(e^{- \b V_0}-1\right)^2.
\end{equation}

Notice the obvious limit $V_0\rightarrow \infty,$ yielding the
familiar Asakura-Oosawa result \cite{AO,AO2}

\begin{equation}\label{phiinf}
\phi^i(\RR_a-\RR_b,\infty)    = -\frac{\rho_0 A_0(\RR_a-\RR_b)}{\b}.
\end{equation}

This problem can also be treated at various orders of response
theory, using, e.g., $\cu(\kk,\kk')=-\b \rho_0  \delta_{\kk,-\kk'}.$
The general form for the pair potential, which can be obtained by
carrying the perturbation to any order, can be expressed as

\begin{equation*}
\phi^i_n(\RR_a-\RR_b,V_0)= -f(\b V_0) \frac{\rho_0
A_0(\RR_a-\RR_b)}{\b}.
\end{equation*}

Only the position-independent function $f$ is modified in the
various approximations. The exact (nonperturbative) form for the function $f$ is

$$f(x)=\left(e^{-x}-1\right)^2.$$

 We evaluate the performance of the various orders in response and of the SPM and SSPM by comparing the estimates
 obtained from each method to the exact result for $f(x).$
 Since we know the exact result analytically, we can obtain the $n^{th}$ order response
estimate to $f$, and hence to the pair potential, by a simple Taylor
expansion of $f(x),$ keeping terms up to $(n+1)^{th}$ order in $x=\beta V_0$,
i.e.,

$$f(x)=x^2-x^3+ \frac{7}{12}x^4+\cdots$$

\begin{figure}
\scalebox{.4}{\includegraphics{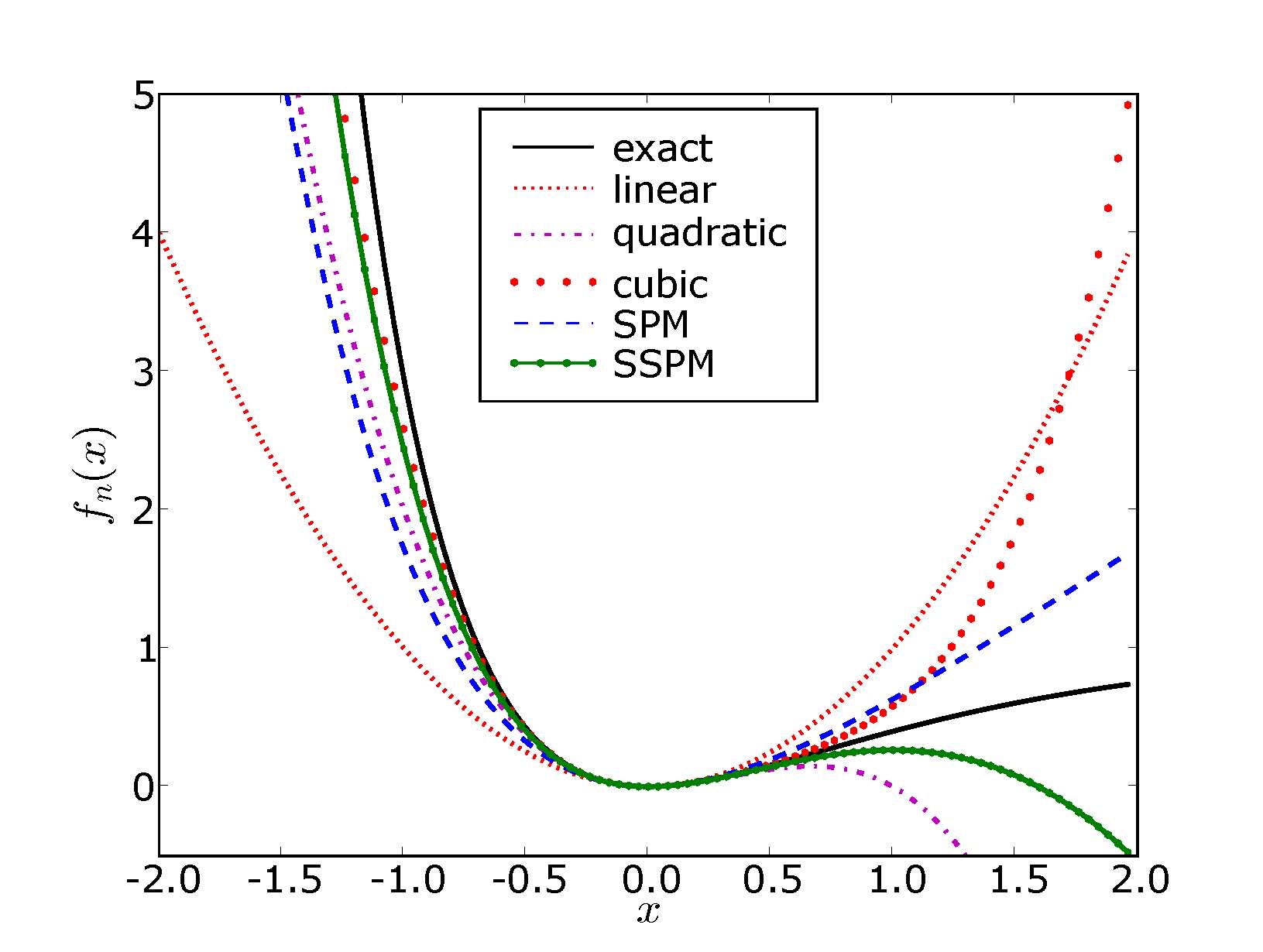}}
\scalebox{.4}{\includegraphics{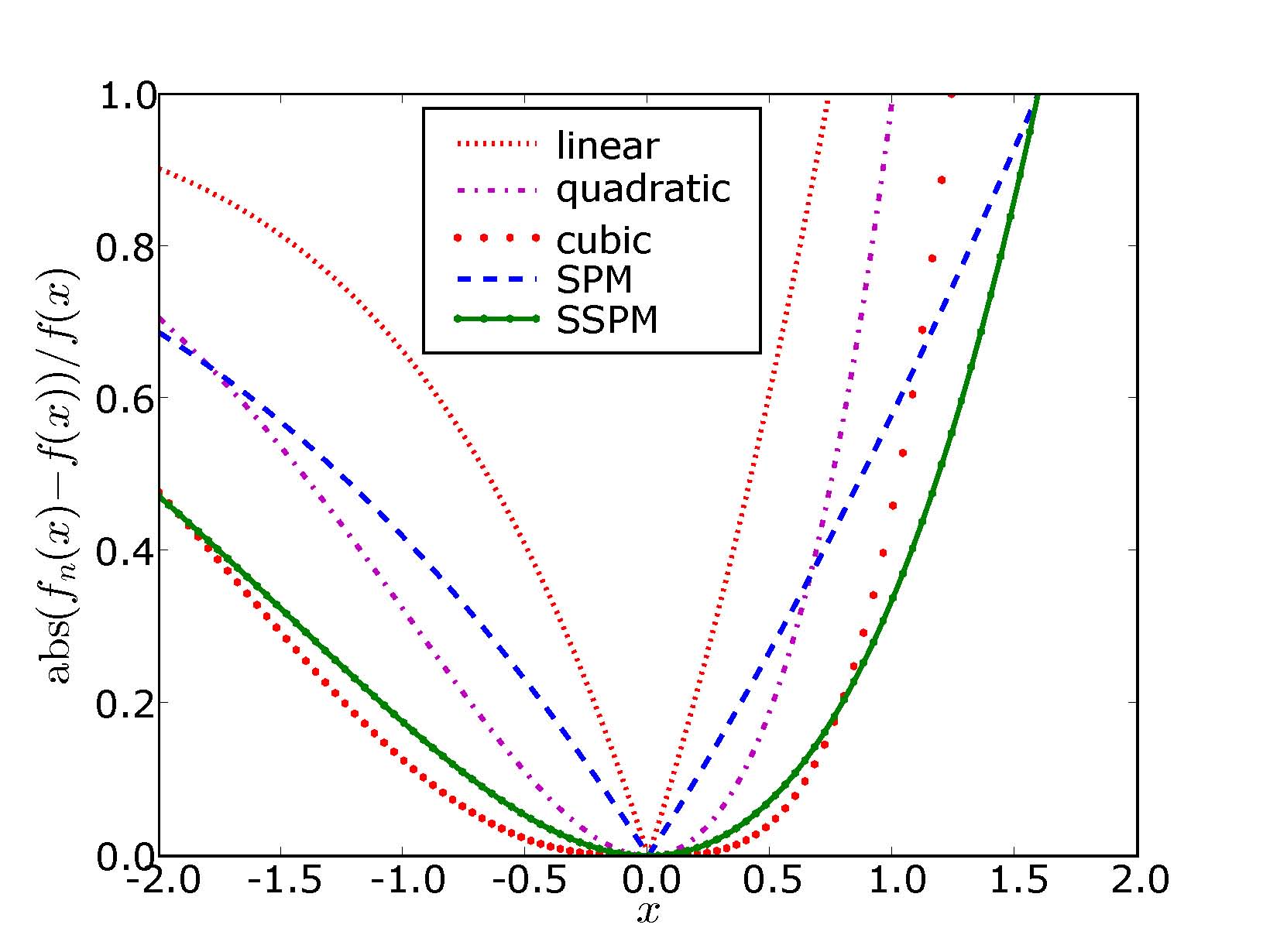}} \caption{(color
online)(upper figure) Dependence of the effective pair potential
between two repulsive colloidal particles on the interaction
strength parameter $x=\beta V_0$: exact result versus the SPM and
SSPM, and perturbations at various orders. (lower figure) Relative
errors $| f_n(x)-f(x)|/f(x)$.}\label{Boltz}
\end{figure}

Expansions up to cubic order are shown on Figure \ref{Boltz},
together with the SPM result [equation \eqref{flin}]

$$f_{SPM}(x)=\left(1-e^{-x}\right)x=x^2-\frac{x^3}{2}+\frac{x^4}{6}+\cdots,$$

\noindent and the SSPM result [equation \eqref{resum}]

$$f_{SSPM}(x)=2\left(x(1-e^{-x})-\frac{x^2}{2}\right)=x^2-x^3+\frac{x^4}{3}+\cdots$$

Negative values of $x$ are included in Figure \ref{Boltz} for
illustrative purposes. One can again observe that because of the
additional terms it includes, the SSPM is more accurate than linear
and quadratic response for all $V_0$ and more accurate than the SPM
for $x<1.59.$
 Cubic response, on the other hand, is closer to the
exact result than the SSPM for $-1.93<x<0.78,$ which is expected
since the SSPM does not include all cubic terms. For larger $|x|,$
though, the higher-order terms play a more important role and the
SSPM is more accurate than cubic response.

This example is also instructive in that it allows us to study
directly the convergence of response theory. Note that for $V_0<0$
the convergence is monotonous, while for $V_0>0$ there is a more
complex, alternating approach to the exact result. This can be
traced back to the fact that changing the overall sign of the
perturbing potential results in changing the sign of odd orders in
response, without affecting the even orders. Thus, if a potential
exhibits monotonous convergence, its additive inverse exhibits
alternating convergence. We might therefore expect, for example,
that effective interactions with protons and antiprotons in an
electron gas will have completely different convergence behaviors.
If this classical example is to be representative, protons would
then exhibit uniform convergence, while antiprotons should exhibit
alternating convergence.

Note, finally, that in the problem at hand, response theory
converges even for arbitrarily large $V_0$ and $D+\delta.$ Since for
repulsive spheres the functions $f(\b V)$ approaches a constant
value exponentially, it is even possible, in this case, to obtain a
quantitative value for the limit $V_0\rightarrow\infty$ by keeping
only a finite number of perturbation terms.

\subsection{The hydrogen molecule and connections with the Heitler-London approach}
\label{Hydrogen} We move on to the more realistic system composed of
two initially bound proton-electron systems immersed in a uniform,
neutral, and \textit{interacting} jellium
\cite{PhysRevB.20.446,PhysRevB.60.2074,PhysRevB.40.1993} at a
temperature much lower than the Fermi temperature. This
hydrogen-in-jellium problem can be linked to real systems in two
ways. First, it can be seen as a first step in a formal expansion
including three- and many-center terms which, if carried to all
orders, should yield the exact total energy of the system, within
the adiabatic approximation. In can also be used within an
effective-medium approach\cite{JNP} in an attempt to take into
account the many-ion effects in an approximate way. In both cases
the pair potentials can be used to derive phonon spectra. In
particular the pair potentials obtained from quadratic response were
used to predict the infrared and Raman vibron frequencies
\cite{NBBA}.

To establish a basis of comparison, we first obtain an estimate of
the importance of nonlinear corrections to the one-atom density,
$\dr_a,$ using the ab initio program VASP \cite{VASP1,VASP2}. This
allows us to estimate the effect on the pair potentials of the
higher-order terms in the determination of $\dr_a.$

 To obtain this, we used a cubic cell of side $13.5 a_0$ containing 74
electrons (for $r_s=2$), together with a $6\times6\times6$ $k$-space
grid and a standard projector augmented wave (PAW) pseudopotential
for hydrogen \cite{kresse1999upp,blochl1994paw} with cutoffs up to
$450$ eV. The generalized gradient approximation (GGA) to the
exchange-correlation potential, as parameterized by Perdew and Wang
\cite{perdew1991ess}, was used, together with the Methfessel-Paxton
smearing \cite{methfessel1989hps}, with a smearing temperature of
$\sig=0.2$ eV. Since it was shown \cite{BA} that the proton-proton
pair potential does not depend strongly on the choice of a
pseudopotential, cutoff energy, and exchange-correlation functional,
our choice of parameters should yield sufficient precision for our
purposes, even considering possible short-range distortions due to
core overlap.

\begin{figure}
\scalebox{.5}{\includegraphics{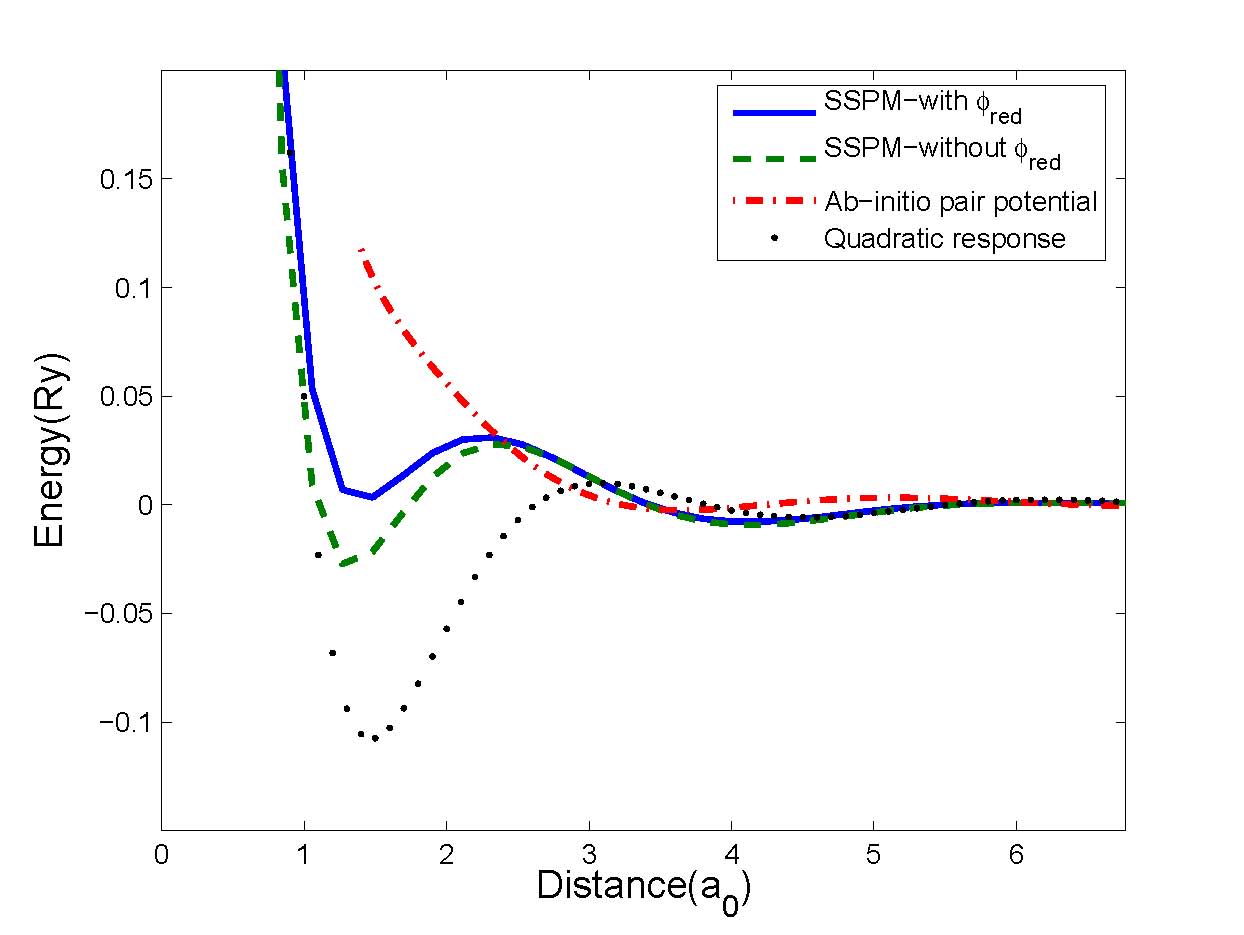}} \caption{(color online)
Comparison of the SSPM results with and without contributions from
equation \eqref{elel} to quadratic response and ab initio results
(from Bonev and Ashcroft\cite{BA}) for proton-proton pair potentials
at $r_s=2$.\label{VASPfig}}
\end{figure}

 The SSPM pair potentials thus obtained are shown in Figure
 \ref{VASPfig}, and are compared with the VASP results for the pair
 potentials from Bonev and Ashcroft\cite{BA}, the quadratic response results, and the SSPM
 pair potentials corrected by the inclusion of equation
 \eqref{elel} to lowest order in the perturbing potential.

\begin{figure}

\scalebox{.4}{\includegraphics{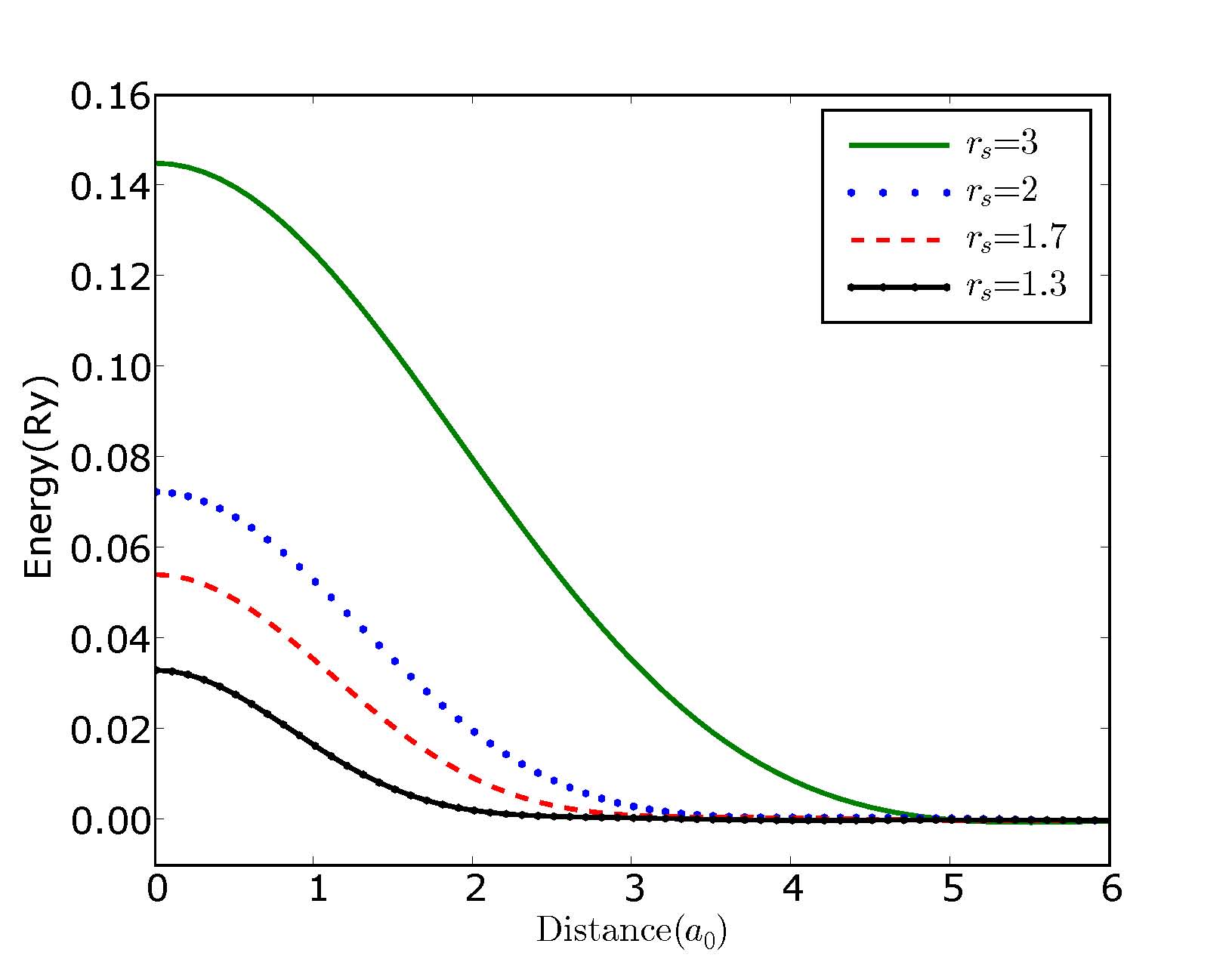}} \caption{(color online)
Contribution of the diagram from Figure \ref{diaglunettes}(a) to the
proton-proton pair potential for various values of
$r_s$\label{lunettes}.}

\end{figure}

 To obtain the latter, we start with \eqref{elel} and use
$$\delta\rho^{NL}_a(\kk)\simeq\frac{1}{\O}\sum_{\kk',\kk''}
\cd(-\kk,\kk',\kk'') V_{ext} (\kk') V_{ext}(\kk'').$$

\begin{figure}
\scalebox{.33}{\includegraphics{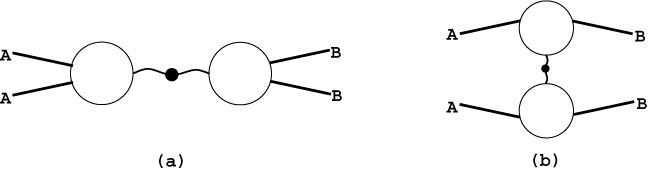}} \caption{(a)
Lowest-order diagram contributing to the interaction \eqref{elel}.
(b) Nonreducible diagram equal to diagram (a) in the limit
$R_{ab}\rightarrow 0,$ for identical ions, here protons
(diagrammatic conventions are explained in Appendix I and Figure
\ref{exemples}).\label{diaglunettes}}
\end{figure}

This corresponds to the energy diagram shown as Figure
\ref{diaglunettes}(a).

We use the second-order noninteracting response function of the
homogeneous electron gas, $\cd_0(\kk,\kk',\kk'),$ in the form given
by Milchev and Pickenhain \cite{MP}. The interacting response
function is then written as

$$\cd(\kk,\kk',\kk')=\frac{\cd_0(\kk,\kk',\kk')}{\eps(\kk)\eps(\kk')\eps(\kk'')},$$

\noindent which corresponds to the random phase approximation. This
approximation is expected to improve as the density is increased
(see, e.g., Pines and Nozi\`eres\cite{PN}), and should be sufficient
to give us information on the general behavior of the potential.
Finally we choose the Vashishta-Singwi form \cite{VS} for the local
field correction $G(\kk)$.

Under these assumptions we obtain the contribution to the pair
potentials displayed in Figures \ref{VASPfig} and \ref{lunettes}.

 We see that the inclusion of the potential $\phi_{red}$ from equation \eqref{elel},
 and therefore of the density-density interaction $\tilde
v(q)=\eps(q) \left[v_c(q)+\mu_1(q)\right],$ leads to a contribution
to the pair potential that is repulsive at typical proton-proton
separation. This behavior largely arises from the Coulomb repulsion
term $v_c.$ This contribution therefore explains part, but not all,
of the discrepancy observed between the pair potentials obtained
from quadratic response \cite{NBBA} and ab initio methods \cite{BA}.
In particular, all perturbation-based methods yield a local minimum
around $R=1.4 a_0,$ whereas no such minimum is observed in the VASP
pair potential at the considered density; such a minimum appears for
VASP pair potentials only at lower densities, corresponding to
$r_s\gtrsim3$ (see Bonev and Ashcroft \cite{BA}). The consequences
of the disappearance of this local minimum at higher density for the
stability of the hydrogen molecule (or crystal) has also been
discussed in Nagao \textit{et al} \cite{NBBA} and D\'iez Mui\~no and
Salin\cite{PhysRevB.40.1993}.

 Note that for each diagram
of the form shown in Figure \ref{diaglunettes}(a) there are two
similar diagrams of the form shown in Figure \ref{diaglunettes}(b)
which are not reducible and hence not contained in \eqref{elel}. In
the limiting case where the interparticle distance tends to zero,
though, diagrams \ref{diaglunettes}(a) and \ref{diaglunettes}(b)
should provide the same contribution. Therefore diagrams of the form
\ref{diaglunettes}(b) should also have an overall repulsive
behavior, contributing to further reduce the discrepancy between
perturbative and VASP results.

\begin{figure}
\scalebox{.4}{\includegraphics{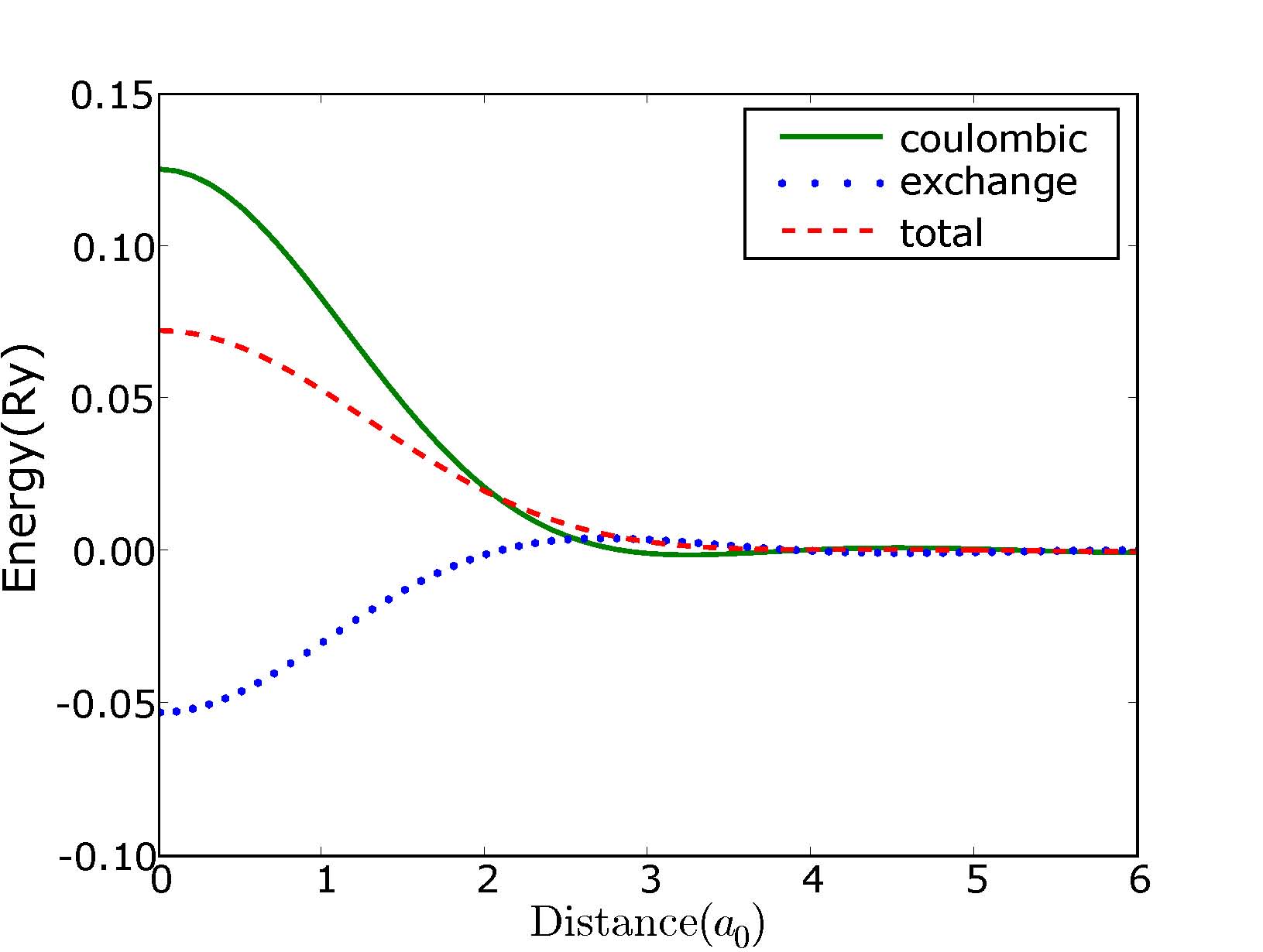}} \caption{(color online)
Contributions of the Coulombic and exchange parts to the effective
electron-electron potential $\tilde v,$ for $r_s=2.$
 \label{mucoul}}
\end{figure}

Note also that if we consider separately the contributions of
$\tilde v_c(q)=\eps(q) v_c(q)$ and $\tilde \mu_1=\eps(q) \mu_1(q),$
we find that the contribution of $\tilde\mu_1$ is mostly attractive,
as can be seen in Figure \ref{mucoul} for $r_s=2$. We also observe
that terms arising from equation \eqref{elel} exhibit very weak
Friedel oscillations. This distinguishes these terms from the other
contributions calculated here or ab initio pair potentials \cite{BA}
(compare Figures \ref{lunettes} and \ref{VASPfig}), and we
conjecture that this arises from the reducibility of the energy
diagram.

 We draw the reader's attention to the close similarity of the pair potential of equation
\eqref{final} with corresponding terms in the Heitler-London (HL)
picture of the isolated hydrogen molecule. We can identify in both
methods (i) the ion-ion repulsions, (ii) the attractive interactions
between ion $b$ ($a,$ respectively) with the density induced by $a$
($b,$ respectively), (iii) the Coulombic repulsion between the
one-atom electronic densities, and (iv) an attractive exchange
contribution from the electrons. The fourth term in \eqref{final},
which has no equivalent in the HL picture, goes to zero in the low
density, free molecule case. It is interesting that two such
different approaches, one being variational in essence and the other
one perturbational, yield such similar results.

Some features of the HL pictures cannot be observed in the SSPM,
though, because of the nonzero average electronic density considered
in the SSPM. For example, while the pair potential between protons
immersed in spin-polarized or spin-unpolarized electrons differed
only by the exchange term in HL, this is no longer the case in the
SSPM approach. The densities $\dr_i$ induced by single protons are
indeed quite different for spin-polarized and spin-unpolarized
electrons.

\section{Conclusion}

 To obtain effective interactions between particles immersed in a well understood
system, such as ions immersed in a uniform jellium, the traditional
perturbation approach treats all immersed particles as a single
perturbation. One finds the free energy of the perturbed system as a
function of the total perturbation potential. The total potential is
then separated in the sum of its constituents, which allows a
natural separation of the effective potential in pair-, triplet-,
and many-body potentials.

Here we have discussed two alternative approaches. In the first
approach, we start by immersing a single particle in the well
understood system and then calculate the response functions of the
new, but perturbed system. Since these response functions are
directly related to the electronic density or density-density
correlation functions, they can be obtained by a variety of
techniques, including perturbation theory, simulations, and density
functional theory and even deduced from experiment. The remaining
particles are then treated as a further perturbation of this already
perturbed system.

Such a two-step process is ideal when only one perturbation is
strong \cite{pscp}, but it is intrinsically asymmetrical and not
ideal when both perturbations are large enough to induce nonlinear
effects. We have suggested a way to improve and symmetrize this
procedure which allows one to treat an increased number of
perturbation terms with little additional effort. More specifically,
by using results of the asymmetrical approach that are exact to
linear order in the perturbation potential, it is possible to
construct a symmetrized result exact to quadratic order and
including additional higher-order terms as well. We applied this
method to two simple noninteracting test systems and found that it
improves upon standard linear and quadratic response, as expected.

More importantly, it was shown that this simple method could be
naturally refined by the inclusion of higher-order terms describing,
in particular, electron-electron interactions. These higher-order
terms also have intuitive physical meaning, and we used this to
argue that the standard termination of the perturbation series at a
given order is not the best strategy. The inclusion of the
higher-order terms was indeed shown to improve considerably the
agreement between the perturbation and density functional theory
approaches in the problem of proton-proton pair potentials.

Even though we considered here effective interactions between
identical particles only, the SSP method is particularly well suited
to the description of systems with multiple (say $N$) types of
particles. It indeed reduces the computational difficulty from the
determination of $N^2/2$ pair potentials to that of finding $N$
(typically symmetric) induced densities, from which the pair
potentials can be obtained in a straightforward manner. It can also
be generalized to many-center interactions and to magnetic
perturbations (see Appendix II).

Finally, a striking similarity is found between terms in the
pair potentials arising from this approach and from
the Heitler-London variational approach for diatomic molecules,
leading to a possible natural generalization of the Heitler-London
approach to metallic systems. This similarity could be explored
further by comparing the pair potential for a pair of atoms in a
jellium, as derived here, to a pair of atoms in a Wigner-Seitz
spherical cell ensuring the same average density, which could be
treated within a Heitler-London-like approach.

A more detailed comparison of the SSPM and ab initio pair potentials
for hydrogen and other materials, especially at densities higher
than those considered here, would be a logical next step to this
work, as would be a more detailed treatment of the many-body
interactions, both from the ab initio and SSPM perspectives.

\begin{acknowledgements}

We would like to thank Bruno Rousseau for many enlightening
discussions and also Stanimir Bonev, Ji Feng, James Porter and David
Roundy. This work was supported by a ES-B NSERC fellowship, and by
NSF grant number DMR-0601461.

\end{acknowledgements}

\section*{Appendix I: Derivation of equation \eqref{elel}}

 We first use the Hohenberg-Kohn-Sham approach
\cite{HK,KS} (and the finite-temperature extension due to Mermin
\cite{Mer}) to write the induced density in terms of the response
functions of a noninteracting system in a modified external
potential (see, e.g., Lundqvist and March\cite{LMp124}). We write
the density as

\begin{equation}\label{KS}
\begin{split}
\delta \rho(\kk)&=\sum_{\kk'}\cu_0(-\kk,\kk')
\Gamma(\kk')\\&+\frac{1}{\O}\sum_{\kk',\kk''}\cd_0(-\kk,\kk',\kk'')\Gamma(\kk')\Gamma(\kk'')+\cdots,
\end{split}
\end{equation}

\noindent where $\chi^{(n)}_0$ are the response functions of the
noninteracting system and $\Gamma$ is the \emph{effective}
perturbation potential. We write $\Gamma$ as

\begin{equation}\label{gamma}
    \Gamma(\kk,\lam)=\lam V(\kk)+c(\kk,[\rho])-c(\kk,[\rho_0]),
\end{equation}

\noindent where $\b c(\kk,[\rho])$ is the first order direct
correlation function.

The parameter $\lam$ is introduced to keep track of the order of the
expansion. For example, we can write

\begin{equation}\label{gamma2}
    \Gamma(\kk)=\lam \Gamma^{(1)}(\kk)+\lam^2
    \Gamma^{(2)}(\kk)+\cdots
\end{equation}

The variation $\delta
c(\kk,\left[\rho\right])=c(\kk,[\rho])-c(\kk,[\rho_0])$ can in turn
be written as a functional expansion:

\begin{equation}\label{muexp}
\begin{split}
    \delta c(\kk)=& \sum_{\kk'}c_0^{(1)}(-\kk,\kk')
    \dr(\kk')\\&+\frac{1}{\O}\sum_{\kk',\kk''}c_0^{(2)}(-\kk,\kk',\kk'')\dr(\kk')\dr(\kk'')+\cdots
\end{split}
\end{equation}

Note that here $\b c_0^{(i)}$ is the $(i+1)^{th}$ direct correlation
function of the unperturbed system. This slightly unusual notation
is chosen to emphasize the similarity between equations \eqref{KS}
and \eqref{muexp}.

Finally the induced density can also be expressed in powers of the
external potential:

\begin{equation}\label{rho}
\delta\rho(\kk)=\lam\delta\rho^{(1)}(\kk)+\lam^{2}
\delta\rho^{(2)}(\kk)+\cdots
\end{equation}

The variation in the free energy is obtained as before through
equation \eqref{fgen}, which we now write as

\begin{equation}\label{fen}
\begin{split}
\Delta F&=\frac{1}{\O}\sum_{\kk,n} \frac{\rho^{(n)}(\kk)
V_{ext}(-\kk)}{n+1}\\&=\Delta F_1+\frac{1}{\O}\sum_{\kk,n}
\frac{\dr^{(n)}(\kk) V_{ext}(-\kk)}{n+1},
\end{split}
\end{equation}

\noindent where $\Delta F_{1}$ does not depend on the relative
positions of the perturbation sources and will therefore not
contribute to the pair potentials.

\begin{figure*}
\scalebox{.5}{\includegraphics{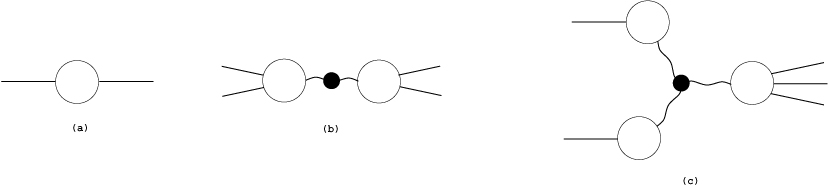}}
\caption{Some typical diagrams from the free energy expansion
contributing to the pair potential. Loops with $n+1$ legs represent
response function of order $n.$ The correlation function $c_0^{(n)}$
is represented by a solid circle surrounded by $n+1$ wavy lines and
the external potential by solid lines.\label{exemples}. Diagram (a)
is a quadratic response contribution. Diagrams (b) and (c) represent
some third-order contributions taken into account by equations
\eqref{resum} and \eqref{elel}, respectively. Diagrams (d), (e), and
(f) are examples of cubic-response contributions which are
\emph{not} taken into account in the approach presented here. }
\end{figure*}

Using equations \eqref{KS}, \eqref{gamma}, and \eqref{muexp}, we can
represent each term in \eqref{fen} by a diagram using three types of
building blocks: namely $\chi_0^{(n)}$ and $c_0^{(n)}$ for $n\geq1,$
and $V_{ext}.$ A few examples are shown in Figure \ref{exemples}.

Each $V_{ext}$ is connected to a single $\chi_0^{(n)}.$ Each
$c_0^{(n)}$ is connected to $n+1$ different $\chi_0^{(n)}.$ Each
$\chi_0^{(n)}$ is connected to $n+1$ blocks that can be either
$V_{ext},$ or a $c_0^{(n)}.$
 Conversely, every treelike diagram
obeying these rules corresponds to a term in equation \eqref{fen}.
Note that since $\dr(\qq)$ obeys the same rules, it can be
represented by the same diagrams, with only one $V_{ext}$ removed.

We call reducible those diagrams in \eqref{fen} that can be
separated, by the cutting of a single $c_0^{(1)}$ line, into two
diagrams that depend exclusively on $V_a$ or on $V_b,$ respectively,
and that are linear in neither $V_a$ nor $V_b.$ In Figure
\ref{exemples}, diagram $(c)$ is reducible, while the others are
not.

We want to show that the set of such diagrams is equivalent to those
described by equation \eqref{elel}. First we will show that there is
a one-to-one correspondence between reducible diagrams in
\eqref{fen} and diagrams in \eqref{elel}. Then we will show that the
prefactors also agree.

\subsection{Diagrammatic equivalence}

Consider a reducible diagram $D$ contributing to the pair potential.
Then consider the set $S$ of all diagrams that are different from
$D$ only by the number of separating interaction lines. The
separating interaction lines can only be connected by $\chi_0^{(1)}$
loops. By summing over the different numbers of $\chi_0^{(1)}$ loops
and the momenta associated with these loops, we can therefore obtain
a dressed propagator in terms of a dielectric function,

\begin{figure*}
\scalebox{.46}{\includegraphics{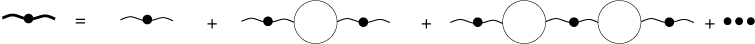}}

\caption{Diagrammatic expansion for the dressed $\tilde c_0^{(1)}$
in terms of the bare $c_0^{(1)}$ and linear response functions.
\label{dressed}}
\end{figure*}

\begin{equation}\label{eps}
 \tilde
 c_0^{(1)}(\kk,\kk')=\frac{c_0^{(1)}(\kk,\kk')}{\eps(\kk,\kk')},
\end{equation}

\noindent which takes a simple form in the diagrammatic language
(see Figure \ref{dressed}).

All diagrams in $S$ are therefore included in $\sum_{\{\kk\}}
D_1(\{\kk\})/\eps(\kk_i,\kk_j)$ where $D_1(\{\kk\})$ is the value,
\emph{before} the summation over the momenta $\{\kk\}$, of the
diagram in $S$ with a single separating $c_0^{(1)}$ propagator.
$\kk_i$ and $\kk_j$ are the momenta of the separating propagator.

Since we can construct diagrams contributing to $\dr(\kk)$ with the
same building blocks and the same construction rules as for diagrams
contributing to the pair potential, it is easy to find an expression
involving only $\dr_i$ and $c_0^{(1)}$ that includes all reducible
diagrams contributing to the pair potential. An example of such an
expression is

\begin{equation*}
   \frac{1}{\O} \sum_{\kk,\kk'} \delta\rho^{NL}_a(\kk,\RR_a) c_0^{(1)}(\kk,\kk')
    \delta\rho^{NL}_b(\kk',\RR_b).
\end{equation*}

 Since both $\dr_a(\kk)$ and $\dr_{b}(\kk)$ have an arbitrary number of
$\cu_0$ loops on the leg with momenta $\kk$, though, this expression
amounts to screening the $c_0^{(1)}$ interaction twice. To take care
of this, we can write instead
\begin{widetext}
\begin{equation}\label{elelinh}
\begin{split}
    \phi_{red}(\RR_{a},\RR_b)=\frac{1}{\O}\sum_{\kk,\kk'} \delta\rho^{NLU}_a(\kk,\RR_a) \frac{c_0^{(1)}(\kk,\kk')}{\eps(\kk,\kk')}
    \delta\rho^{NLU}_b(\kk',\RR_b),
\end{split}
\end{equation}
\end{widetext}
\noindent where $\delta\rho^{NLU}_a(\kk,\RR_a)$ is defined as the
sum of all diagrams in  $\delta\rho^{NL}_a(\kk,\RR_a)$ without $\cu$
loops on the leg with the $\kk$ momentum (U stands for unscreened).
To each reducible diagram contributing to the pair potential
corresponds a diagram in \eqref{elelinh}, and vice-versa.

For homogeneous systems, we have

\begin{equation*}\label{inhomohomo}
    \begin{split}
\cu(\kk,\kk')&=\cu(k')\del_{\kk,-\kk'},\\
c_0^{(1)}(\kk,\kk')&=c_0^{(1)}(k')\del_{\kk,-\kk'},\\
 \frac{1}{\eps(\kk,\kk')}=
\frac{1}{\eps(k)}&=\sum_i(c_0^{(1)}(k)\chi_0^{(1)}(k))^i\\&= \frac{1}{(1-c_0^{(1)}(k)\chi_0^{(1)}(k))},\\
\end{split}
\end{equation*}

\noindent and
\begin{equation*}
 \del\rho_{a,b}(\kk,\RR_{a,b})=\rho_{a,b}(\kk)e^{i \kk\cdot\RR_{a,b}},
\end{equation*}
\noindent so that \eqref{elelinh} simplifies to

\begin{equation*}\label{elelhomo}
    \phi_{red}(\RR_{ab})=\frac{1}{\O} \sum_{\kk} \delta\rho^{NL}_a(\kk) c_0^{(1)}(k) \eps(k)
\delta\rho^{NL}_b(-\kk)e^{i \kk \cdot\RR_{ab}},
\end{equation*}

\noindent which is precisely equation \eqref{elel}; we have shown
that equation \eqref{elel} contains exactly the reducible diagrams
from equation \eqref{fen}. Now we need to show that the prefactors
of these diagrams also agree.

\subsection{Prefactors and symmetries}
 A diagram $D$ is said to possess a symmetry of order $m$ if it is possible to cut $m$ legs from a given
$\chi_0$ or $c_0$ and obtain $m$ identical cut-down parts. Each such
symmetry contributes a factor $1/m!$ to the total prefactor of the
diagram, as compared to an equivalent asymmetrical diagram.

 In order to obtain this "equivalent asymmetrical diagram", we replace $V_{ext}$ by $\tilde
 V_{ext}=\sum_{j=1}^{n+1}
 v_j$ in the original problem, where $n+1$ is the number of external legs of diagram $D.$
 We then consider the diagram $\tilde D$ that is identical to $D,$
apart from its external legs
 which are all connected to different $v_j.$ By construction, diagram $\tilde D$ can have
no symmetry.

Contributions to diagram $\tilde D$ can be obtained, in equation
\eqref{fen}, by replacing $V_{ext}$ by any of the $v_j$ and by
calculating the prefactor $P_j$ of the density diagram obtained by
removing $v_j$ from $\tilde D$. The prefactor $P_{\tilde D}$ of
$\tilde D$ is therefore

\begin{equation}\label{sumloops}
   P_{\tilde D}= \frac{1}{n+1}\sum_{\mbox{$j$}:\substack{\mbox{external}\\\mbox{leg}}}P_j,\\
    \end{equation}

Since there is no symmetry in the diagram, it is relatively
straightforward to find its prefactor: each $\chi^{(i)}_0$ or
$c^{(i)}_0$ contributes $i!$ to the global prefactor. Therefore all
the $P_j$ are equal, the $(n+1)$ factors cancel out, and we simply
get $P_{\tilde D}=P_j.$ By a similar argument, we obtain
$$P_{\tilde{D}}=P_j= P_{\tilde D_a} P_{\tilde D_b},$$ \noindent
where $\tilde D_a$ and $\tilde D_b$ are the separated diagrams.
Asymmetric reducible diagrams therefore have the same prefactor in
\eqref{fen} and \eqref{elel}.

This result can be extended to symmetrical diagrams in a
straightforward manner: if the two identical diagrams obtained by
cutting two legs from a given $\chi$ or $\mu$ in a symmetrical
reducible diagram, they cannot contain the separating interaction
line. Therefore all symmetries are contained within the separated
diagrams. Since each symmetry contributes a factor $1/m!$ to the
global prefactor, $m$ being the number of identical branches
involved in this symmetry, the symmetry contributions to
\eqref{elel} and \eqref{fen} are the same.

We have therefore derived the result that there is a one-to-one
correspondence between reducible diagrams contributing to the pair
potential [in equation \eqref{fen}], and diagrams contained in
equation \eqref{elel}. Since the prefactors of these diagrams in
each expression also agree, equation \eqref{elel} allows the
calculation of the pair potential associated with all reducible
diagrams.

\section*{Appendix II: Generalizations}

In this appendix we discuss two possible generalizations of the
methods introduced above: namely, the case of magnetic perturbations
and that of many-center interactions.

\subsection*{A.Magnetic perturbations and the RKKY interaction}

The interaction between magnetic perturbations (e.g., nuclear spins
or magnetic impurity atoms) and an electron gas can be modeled by
the Hamiltonian \cite{Hew}
 $$H_{int,\lam}=\lam J \sum_{i,j}f(\rr_i-\RR_j)\SS^j  \cdot\ssig^i,$$

\noindent where $\rr_i$ and $\ssig^i$ are the position and spin
operators for electron $i$, while $\RR_j$ and $\SS^j$ are the
position and spin of the $j^{th}$ magnetic perturbation source.

 The Hellman-Feynman theorem reads

\begin{equation}\label{HFRKKY}
\begin{split}
F_1-F_0&= \int_0^1 d\lam \langle H_{int,1}\rangle_\lam
\\
&={J} \int_0^1 d\lam \sum_{j}  \int d\rr  f(\rr-\RR_j) \SS^i\cdot
\bf{m}(\rr)_\lam,
\end{split}
\end{equation}

\noindent  where the local magnetization $\bf{m}$ is given by

$$\bf{m}(\rr)_\lam=\left\langle\sum_i \delta(\rr_i-\rr) \ssig^i\right\rangle_\lam.$$

 By analogy with the induced density case, we can describe the perturbation using $$V_j(\rr)=J\sum_i f(\rr-\RR_i)
S^i_j.$$  Assuming a nonmagnetic unperturbed state,

\begin{equation*}
\begin{split}m_i(\rr)_\lam=&\int d\rr \chi_{ij}(\rr,\rr') \lam V_j(\rr')
\\&+\int d\rr'd\rr'' \chi^{(2)}_{ijk}(\rr,\rr',\rr'') \lam^2 V_j(\rr')
V_k(\rr'')+\cdots
\end{split}
\end{equation*}

If we have two magnetic perturbing sources ($a$ and $b),$ we find an
effective interaction of the form

\begin{equation}\label{SSPrkky}
\begin{split}
&\phi(\RR_a,\RR_b,\SS^a,\SS^b)=\int d\rr V^a(\rr)\cdot
\mm^b(\rr,\RR_b,\SS^b)\\&+V^b(\rr)\cdot \mm^a(\rr,\RR_a,\SS^a) -\int
d\rr d\rr' \chi_{ij}(\rr,\rr') V^a_i(\rr) V^b_j(\rr').
\end{split}
\end{equation}

Here, as before, $V^{i}$ describes the perturbation associated with
source $i$ and $\mm^i(\rr,\RR_{i},\SS^a)$ is the magnetization
induced at $\rr$ by the presence of a single perturbation of spin
$S^{i}$ at $R_{i}.$

In the particular case of pointlike magnetic perturbations
[$f(\rr)=\delta(\rr)$], the effective interaction between the
magnetic perturbations takes the simple form

\begin{equation}
\begin{split}
&\phi(\RR_a,\RR_b,\SS^a,\SS^b)= J S^a \cdot
\mm^b(\RR_a,\RR_b,\SS^b)\\&+J S^b\cdot \mm^a(\RR_b,\RR_a,\SS^a)
-J^2\chi_{ij}(\RR_a,\RR_b) S^a_i S^b_j.
\end{split}
\end{equation}

This expression, which goes beyond quadratic response [but does not
include corrections of the form \eqref{elel}], could be used to
improve upon the standard RKKY potential, which takes into account
only the linear order in response \cite{Kit1}.

\subsection*{B.Many-center potentials}

Since the sources of the perturbations in equations \eqref{resum}
and \eqref{elel} have not been specified, a straightforward way of
treating many-particle interactions in this formalism is to take
either or both sources to be an ensemble of particles. This might be
especially appropriate for problems involving the diffusion of well
formed molecules. In the hydrogen problem, this could also be used
to study the molecule-molecule interactions near or beyond the onset
of metallization.

Many-center interactions can be treated in a more symmetric way
through conventional response theory (see, e.g., Porter
\cite{Porter}), but also through a SSPM approach. Many-center
interactions can also be treated in a more symmetric way. The
lowest-order $N$-body term arises at the level of $(N-1)^{th}$ order
response. This term is linear in the potential from all $N$
particles. The next most dominant contributions will be from terms
that are nonlinear in the potential arising from one particle and
linear in that arising from the remaining $N-1.$

We can directly generalize equation \eqref{resum} to an N-body
potential treating all terms linear in $N-1$ particles exactly by
writing
\begin{widetext}
\begin{equation}\label{manybody}
\phi^{(N)}(\{\RR\})\simeq\frac{1}{(N-1)\O} \sum_{j=1}^N \sum_{\kk}
V_j(\kk) \delta \tilde\rho_{\left[j\right]}(-\kk)e^{i \kk\cdot
\RR_j}-\frac{1}{(N-1)\O^{N-1}}\sum_{\{\kk_i\}}
\chi^{(N-1)}(\kk_1,\ldots,\kk_N) \prod_{j=1}^{N} V_{j}(\kk_j)e^{i
\kk_j\cdot \RR_j}.
\end{equation}
\end{widetext}

Here $\delta \rho_{\left[ j\right]}(\kk)$ is the density induced by
all involved particles \emph{except} $j$ and $\delta \tilde
\rho_{\left[ j\right]}(\kk)$ is the component of this density that
depends on the position of all $N-1$ particles involved. The latter
definition is required simply to avoid double counting the
potentials involving less than $N$ particles.

 Equation \eqref{manybody}
provides in principle a way to obtain an expression that is exact up
to $N^{th}$ order, requiring only the explicit knowledge of the
$(N-1)^{th}$ order response function. On the other hand, it requires
knowledge of the density induced by $N$ groups of $N-1$ particles,
which rapidly becomes more difficult when $N>2.$

%\bibliography{../bib/hydrogen}

\begin{thebibliography}{45}
\bibitem[{com({\natexlab{a}})}]{commentaire}
\bibinfo{note}{In the case of neutral systems composed oppositely charged
  particles, such as metals, care should be taken to define the perturbing and
  perturbed systems in such a way that thermodynamic functions such as
  $F_A(\{\RR\},T)$ are well defined. This can be achieved by requiring that
  each physical, charged particle "carries" a uniform compensating charged
  background filling the whole available volume, ensuring a cancellation of
  divergent $\kk=0$ terms. In this case both the perturbed and perturbing
  systems are neutral and thermodynamically well defined, and in the combined
  system the compensating backgrounds cancel out so we recover the physical
  system. In the case of the electron gas the perturbed system is simply
  jellium}.
\bibitem[{com({\natexlab{b}})}]{commentHL}
\bibinfo{note}{It is assumed here that the free energy varies continuously
  while $\lam$ goes from $0$ to $1$}.
\end{thebibliography}
\bibstyle{apsrev}

\end{document}